\documentclass[prc,showpacs,showkeys,superscriptaddress,nofootinbib,floatfix,onecolumn]{revtex4}
\usepackage[cp1251]{inputenc}
\usepackage[T2A]{fontenc}
\usepackage[english]{babel}
\usepackage{color}
\usepackage{amsfonts}
\usepackage{amsbsy}
\usepackage{mathrsfs}
\usepackage{graphicx}
\usepackage{subfigure}
\newcommand{\eeq}{\end{equation}}
\newcommand{\bea}{\begin{eqnarray}}
\newcommand{\eea}{\end{eqnarray}}

\def\lsim{\mathrel{\rlap{
\lower4pt\hbox{\hskip-3pt$\sim$}}
    \raise1pt\hbox{$<$}}}     
\def\gsim{\mathrel{\rlap{
\lower4pt\hbox{\hskip-3pt$\sim$}}
    \raise1pt\hbox{$>$}}}     

\begin{document}
\title{Fermionic path integral Monte Carlo results for the uniform electron gas at finite temperature}
\author{V.S.~Filinov}
\author{V.E.~Fortov}
\affiliation{Institute for High Energy Density, Russian Academy of Sciences,
Izhorskaya 13/19, Moscow 127412, Russia
}
\author{M.~Bonitz}
\author{Zh. Moldabekov}
\affiliation{Institut f\"ur Theoretische Physik und Astrophysik, Christian-Albrechts-Universit{\"a}t zu Kiel, 
Leibnizstrasse 15, 24098 Kiel, Germany
}
\begin{abstract}
The uniform electron gas (UEG) at finite temperature has recently attracted substantial interest due to the epxerimental progress in the field of warm dense matter. 
To explain the experimental data accurate theoretical models for high density plasmas are needed which crucially depend on the quality of the thermodynamic properties of the quantum degenerate correlated electrons. Recent fixed node path integral Monte Carlo (RPIMC) data are the most accurate for the UEG at finite temperature, but they become questionable at high degeneracy when the Brueckner parameter $r_s$ becomes smaller than $1$. Here we present new improved direct fermionic PIMC simulations that are exptected to be more accurate than RPIMC at high densities.
\end{abstract}
\maketitle
%
\section{Introduction} 
In recent years the interest in high density plasmas has increased steadily. Examples are dense plasmas in the interior of the giant planets \cite{knudson_12} and compact stars as well as highly compressed laboratory plasmas, such as laser plasmas or inertial confinement fusion plasmas \cite{lindl_04}, for a recent experimental study see \cite{continuum14}. In these systems often the electrons are quantum degenerate and weakly or moderately coupled whereas the ions are classical and moderately and sometimes strongly coupled. For both components it is crucial to include finite temperature effects which poses particular challenges for theory \cite{green-book, ktnp04} and computer simulations, e.g. \cite{holst08, clerouin06} and references therein.

The properties of the electron gas are a crucial ingredient to correctly describe dense plasmas as well as the electron gas in metals. Accurate data for the electron gas at zero temperature have been provided long ago by quantum Monte Carlo simulations \cite{ceperley_alder} and have been massively used in density functional calculations.
In the mean time new and improved ground state data have appeared, the most accurate ones, apparantly, being the configuration interaction Monte Carlo data of Refs.~\cite{alawi_prb12,alawi_jcp12}. However, these are all restricted to zero temperature and not applicable to highly excited systems such as warm dense matter.

The extension of ab initio simulations to finite temperature is possible using the path integral Monte Carlo (PIMC) technique, 
e,g, \cite{feynman-hibbs, binder96, ceperley95, ceperley95rmp, zamalin,numbook},
and many results have been obtained for fermions in the recent two decades, including correlated electrons in quantum dots, e.g. \cite{egger_blocking, afilinov-etal.01prl} or dense plasmas, e.g. \cite{mil-pol,militzer_06,filinov-etal.00jetpl}.

Only recently PIMC simulations were applied to the electron gas at finite temperature.
Brown {\em et al.} presented restricted path integral Monte Carl (RPIMC) results that used the fixed node approximation  \cite{brown13} that cover a broad parameter range of the homogeneous electron gas. Subsequently, semi-analytical fits were presented that intended to combine the PIMC results with the known analytical limits \cite{karasiev14}. 
A key problem of these data is that the accuracy of the results at high degeneracy is unknown since the fixed nodes used in the RPIMC simulations carry a systematic error that is difficult to quantify. As a test, Brown et al. also performed fermionic simulations and observed increasing discrepancies at high degeneracy ($r_s \le 1.5$). At the same time they encountered large errors of the fermionic simulations since the average sign was very small, which is a direct manifestation of the fermion sign problem. Thus, the question of reliable high-temperature data for the uniform electron gas at high densities remains.
 
Therefore, the motivation of the present work is to perform new independent first-principle direct fermionic PIMC simulations and compare them to the earlier results. 
To this end we make use of a PIMC approach that was developed several years ago and successfully applied to dense hydrogen \cite{Filinov01, FiBoEbFo01,Bonitz_PRL05}, hydrogen-helium mixtures \cite{filinov_cpp_05, levashov_jpa_06}, to the electron-hole plasmas in semiconductors \cite{eh,filinov_jpa_03} and to the nonideal quark gluon plasma \cite{filinov_cpp12, filinov_prc13}. To simulate the uniform electron gas the code is modified such that the positive charge component is treated as an ideal gas neglecting correlations. Furthermore, to be able to simulate situations of high degeneracy, the treatment of exchange effects is substantially improved. Details of the PIMC scheme are described in Sec.~\ref{theory} and the numerical results are discussed in Sec.~\ref{simulations}.

\section{Fermionic path integral Monte Carlo simulations}\label{theory}
\subsection{Jellium model}

The neutral uniform electron gas (UEG) or jellium is a quantum mechanical model of interacting electrons where the positive charges (e.g. atomic nuclei of a solid) are assumed to form a uniform in space charge background that assures overall charge neutrality but is not treated microscopically. The model allows one to focus on the effects that occur due to the quantum nature of electrons and their repulsive interactions and to treat the electron -- electron interaction rigorously. 
The artificial and structureless background charge interacts electrostaticaly with itself and the electrons. 

The jellium Hamiltonian (for a text book disussion, see Ref.~\cite{mahan-book}) for $N$ electrons confined within a volume of the space $V$ and with electronic density 
$\rho(r)=\sum_{a=1}^{N}\delta(r-r_a)$ and background charge density $n(R)=N/V$ is 
\begin{equation}\label{ham}
\hat{H}=\hat{H}_{el}+\hat{H}_{back}+\hat{H}_{el-back},
\end{equation} 
where $\hat{H}_{el}$  is the electron Hamiltonian consisting of kinetic and electron--electron repulsion terms: 
\begin{equation}\label{el}
\hat{H}_{el}=\sum_{a=1}^{N}\frac{\hat{p}_a^2}{2m}+\sum_{a<b}^{N}\frac{e^2}{|{\hat r}_a-{\hat r}_b|}.
\end{equation} 
$\hat{H}_{back}$ is the Hamiltonian of the positive background charge describing  its electrostatic self--interaction. Its expectation value is 
\begin{equation}\label{bck}
\langle \hat{H}_{back}\rangle =\frac{e^2}{2}\int_V dR\int_V dR'\frac{n(R)n(R')}{|R-R'|}=
\frac{e^2}{2}\frac{N^2}{V^2}\,\int_V dR\int_V dR'\frac{1}{|R-R'|} 
= \frac{N^2}{2V}\lim_{{\bf q}\to 0} v_q,
\end{equation} 
where we introduced the Fourier component of the Coulomb potential, $v_q = \frac{4\pi e^2}{q^2}$.
The contribution of the electrostatic electron--background interaction 
is given by
\begin{equation}\label{e-bck}
\langle \hat{H}_{el-back}\rangle =-\int_V dr\int_V dR\frac{e^2\rho(r)n(R)}{|r-R|}=-e^2\frac{N}{V}\sum_{a=1}^{N}\int_V dR \frac{1}{|r_a-R|}
= - \frac{N^2}{V}\lim_{{\bf q}\to 0} v_q.
\end{equation} 
In a finite system, the results (\ref{bck}) and (\ref{e-bck}) are finite because $q$ approaches a finite minimal value determined by the system volume. 
If, however, the thermodynamic limit is taken, this limit for $q$ becomes zero, as indicated in the formulas and the two results diverge. This 
divergence is exactly canceled by the $q=0$ contribution of the electron--electron interaction which has the form \cite{mahan-book}
\begin{equation}
 \frac{N(N-1)}{2V}\lim_{{\bf q}\to 0} v_q.
\nonumber
\end{equation}
This cancellation is due to the assumption of charge neutrality and allows to neglect the background contribution entirely and just study the electron component, omitting the $q=0$ contribution (i.e. the spatially homogenous part) to the thermodynamic quantities. 
 
In our simulations we employ a finite simulation cell. For this case, again, the background related energies can be calculated. Assuming, for simplicity, a spherical simulation cell of radius $R$ with a particle number chosen such that a given density $n$ is realized, i.e. $N=n \frac{4\pi}{3}R^3$, one readily obtains from standard electrostatics for the background energy (electrostatic selfenergy)
\begin{equation}
 \langle \hat{H}_{back}\rangle(N) = \frac{3}{5}\frac{Q^2}{R(N)}.
\nonumber
\end{equation}
Similarly, the interaction energy of $N$ electrons homogeneously distributed in this sphere with the homogeneous charge background is found to be 
\begin{equation}
 \langle \hat{H}_{el-back}\rangle(N) = -\frac{6}{5}\frac{Q^2}{R(N)} \equiv -2  \langle \hat{H}_{back}\rangle(N),
\nonumber
\end{equation}
and equals exactly minus two times the background energy, independently of the system size.
Both terms diverge with increasing system size. But one readily sees that the sum of both energies, again, is exactly cancelled by the analogous contribution of the electrons which equals $ \langle \hat{H}_{back}\rangle(N)$. This underlines that all electrostatic mean field contributions compensate each other, and the non-vanishing reminder of the interaction energy is just due to electronic correlations i.e. density fluctuations around the uniform density. 

We can now estimate the scaling of the background energy contribution with system size and density. To this end we switch to atomic units introducing the density (Brueckner) parameter, $r_s={a/a_B}$, where the mean interparticle distance is given by $a=\left[\frac{3}{4\pi n} \right]^{1/3}$, $a_B$ denotes the Bohr radius, and energies are given in Hartree with $1{\rm Ha}=\frac{e^2}{4\pi\epsilon_0}\frac{1}{a_B}$. Then we obtain for the background energy per particle:
\begin{equation}
\frac{\langle \hat{H}_{back}\rangle(N)}{N {\rm Ha}}  = \frac{3}{5}\frac{N^{2/3}}{r_s}.
\nonumber
\end{equation}

\subsection{Path integral representation of thermodynamic quantities}
Particle simulations of Coulomb systems face the problem of an unlimited increase of the interation energy at small distances. Quantum simulations such as path integral Monte Carlo avoid the Coulomb divergences in a natural way. This has been demonstrated before for PIMC simulations of two-component electron-ion plasmas  e.g. \cite{Filinov01, FiBoEbFo01} and electron-hole plasmas \cite{filinov_jpa_03}. Here we adapt these simulations to jellium by treating it as the limiting case of the neutral two-component plasma ($N_e=N_p=N$) where, at the end, the background component (the ions) will be treated as non-interacting. 

Let us start from a quantum two-component Coulomb system of electrons and positive charges in equilibrium with the Hamiltonian, 
${\hat H}={\hat K}+{\hat U}^c$,
containing kinetic energy ${\hat K}$ and Coulomb interaction energy
${\hat U}^c = {\hat U}_{pp}^c + {\hat  U}^c_{ee} + {\hat
U}^c_{ep}$ contributions. The thermodynamic properties in the
canoncial ensemble with given temperature $T$ and fixed volume $V$ are fully described by the density operator ${\hat \rho} = e^{-\beta {\hat H}}/Z$, with the partition function (normalization constant)
\begin{equation}\label{q-def}
Z(N_e,N_p,V;\beta) = \frac{1}{N_e!N_p!} \sum_{\sigma}\int\limits_V
dq \,\rho(q, \sigma ;\beta),
\end{equation}
where $\beta=1/k_BT$, and $\rho(q, \sigma ;\beta)$ denotes the diagonal elements of the density matrix in coordinate representation at a given value $\sigma$ of the total spin. In Eq.~(\ref{q-def}), $q=\{q_e,q_p\}$ and $\sigma=\{\sigma_e\}$ are the spatial coordinates and spin degrees of freedom
of the electrons, i.e. $q_a=\{q_{1,a}\ldots q_{l,a}\ldots q_{N_a,a}\}$
and $\sigma_e=\{\sigma_{1,e}\ldots \sigma_{l,e}\ldots \sigma_{N_e,e}\}$.
In order to calculate thermodynamic functions, the logarithm of the partition function has to be differentiated with respect to
thermodynamic variables. For example, for pressure and internal energy it follows
\begin{eqnarray}
\beta p &=& \frac{\partial {\rm ln} Z}{\partial V} = \left[\frac{\alpha}{3V}
\frac{\partial{\rm ln} Z}{\partial \alpha}\right]_{\alpha=1}, \label{p_gen}
\\
\beta E &=& -\beta \frac{\partial {\rm ln} Z}{\partial \beta},
\label{e_gen}
\end{eqnarray}
where $\alpha= L/L_0$ is a length scaling parameter.

The exact density matrix of interacting quantum systems is not known (particularly for low temperatures and high
densities), but can be constructed using a path integral approach~\cite{feynman-hibbs} based on the operator identity,
\begin{equation}
e^{-\beta {\hat H}}= e^{-\Delta \beta {\hat H}}\cdot
e^{-\Delta \beta {\hat H}}\dots  e^{-\Delta \beta {\hat H}}, \quad \Delta \beta = \beta/(n+1)
\end{equation}
that involves $n+1$ identical high-temperature factors with 
 temperature $(n+1)T$, which allows us to
rewrite the integral in Eq.~(\ref{q-def})
\begin{eqnarray}
&&\sum_{\sigma} \int\limits dq^{(0)}\,
\rho(q^{(0)},\sigma;\beta) =
\int\limits  dq^{(0)} \dots
dq^{(n)} \, \rho^{(1)}\cdot\rho^{(2)} \, \dots \rho^{(n)} \times
\nonumber\\
&&\sum_{\sigma}\sum_{P_e} (\pm 1)^{\kappa_{P_e}} \,
{\cal S}(\sigma, {\hat P_e} \sigma_{a}^\prime)\, \times
{\hat P_e} \rho^{(n+1)}\big|_{q^{(n+1)}= q^{(0)}, \sigma'=\sigma}\,.
 \label{rho-pimc}
\end{eqnarray}
The spin gives rise to the spin part of the density matrix (${\cal S}$) with exchange effects accounted for by the permutation
operators  $\hat P_e$ acting on the electron coordinates $q^{(n+1)}$ and spin projections $\sigma'$. The
sum is over all permutations with parity $\kappa_{P_e}$. Equation.~(\ref{rho-pimc}) involves the off-diagonal high-temperature density matrices, 
$\rho^{(l)}\equiv \rho\left(q^{(l-1)},q^{(l)};\Delta\beta\right) =
\langle q^{(l-1)}|e^{-\Delta \beta {\hat H}}|q^{(l)}\rangle$, where $l=1\dots n+1$.
Accordingly, each particle is represented by a set of $n+1$ coordinates 
(``beads''), i.e. the whole configuration of the particles is represented by a $3(N_e+N_p)(n+1)$-dimensional vector
\begin{equation}
\tilde{q}\equiv\{q_{1,e}^{(0)}, \dots q_{1,e}^{(n+1)},
q_{2,e}^{(0)}\ldots q_{2,e}^{(n+1)}, \ldots q_{N_e,e}^{(n+1)};
q_{1,p}^{(0)}\ldots q_{N_p,p}^{(n+1)} \}.
\nonumber 
\end{equation}

To determine the energy  in the path integral representation~(\ref{rho-pimc}) each high-temperature density matrix has to be
differentiated~\cite{FiBoEbFo01,zamalin}
\begin{eqnarray}
\hspace*{-1cm}\beta E &=& - \frac{1}{Z} \int\limits dq^{(0)} \dots
dq^{(n)}\,\sum_{\sigma}\sum_{P_e} (\pm 1)^{\kappa_{P_e}} \,{\cal S}(\sigma, {\hat P_e} \sigma')\,
\nonumber\\
&&\;\times  \sum_{k=1}^{n+1}\rho^{(1)}
\dots
\rho^{(l-1)}
\left[\beta\frac{\partial \rho^{(l)}}{\partial \beta}\right]
\rho^{(l+1)} \, 
\dots \, \rho^{(n)}{\hat P_e} \rho^{(n+1)}
\bigg|_{q^{(n+1)}=q^{(0)},\, \sigma'=\sigma}\,. \label{e-pimc}
\end{eqnarray}
It is straightforward to show that the matrix elements $\rho^{(l)}$ can be rewritten as
\begin{eqnarray}
\hspace*{-1cm}\rho^{(l)}&\equiv&\langle q^{(l-1)}|e^{-\Delta \beta {\hat
H}}|q^{(l)}\rangle =
\int d{\tilde p}^{(l-1)}d{\bar p}^{(l)}\, \langle
q^{(l-1)}|e^{-\Delta \beta {\hat U}^c}|{\tilde p}^{(l-1)}\rangle
\langle {\tilde p}^{(l-1)}|e^{-\Delta \beta {\hat K}} |{\bar
p}^{(l)}\rangle
\, \nonumber\\
&&\hspace*{6cm}\times \langle {\bar p}^{(l)}| \,e^{-\frac{\Delta
\beta^2}{2}[{\hat K},{\hat U}^c]} \, \dots|q^{(l)}\rangle,
\label{rho_ku}
\end{eqnarray}
where ${\tilde p}^{(l-1)}({\bar p}^{(l)})$ are the conjugate variables to
$q^{(l-1)}(q^{(l)})$. To further evaluating the derivatives in Eq.~(\ref{e-pimc}), it is convenient to introduce dimensionless integration variables
$\eta^{(l)}=(\eta_p^{(l)},\eta_e^{(l)})$, where
$\eta_a^{(l)}=\kappa_a(q^{(l)}_a-q^{(l-1)}_a)$, for $l=1,\dots, n$,
and we introduced the dimensional factor $\kappa_a^2\equiv m_a /(2\pi \hbar^2 \Delta  \beta
)=1/\lambda_{\Delta ,a}^2$, for a=e, p. The main advantage is that differentiation of the density matrix now affects only the interaction terms
\begin{eqnarray}
\beta \frac{\partial \rho^{(l)}}{\partial \beta} &=& -\beta
\frac{\partial [\Delta\beta \cdot U^c(X^{(l-1)})]}{\partial \beta}
\rho^{(l)} +\beta {\tilde \rho}_{\beta}^{(l)}, \label{rho-prime}
\end{eqnarray}
where
\begin{equation}
\hspace*{-1cm}{\tilde \rho}_{\beta}^{(l)} = \int d{p}^{(l)}\, \left\langle
X^{(l-1)}\left|e^{-\Delta \beta {\hat U}^c}\right|{p}^{(l)}\right\rangle
e^{-\frac{\langle {p}^{(l)}|{p}^{(l)}\rangle}{4\pi(n+1)}}
\,
\left\langle {p}^{(l)}\left|\frac{\partial}{\partial \beta}
\,e^{-\frac{(\Delta\beta)^2}{2}[{\hat K},{\hat U}^c]} \,
\dots \right|X^{(l)}\right\rangle  \label{rho-tilde}
\end{equation}
with $p_a^{(l)}={\tilde p}_a^{(l)}/(\kappa_a\hbar)$,
$p{(l)}\equiv \left(p_p^{(l)},p_e^{(l)}\right)$, in accordance with Eq.~(\ref{rho_ku}).
Furthermore
$X^{(0)}\equiv (\kappa_p q_p^{(0)},\kappa_e q_e^{(0)})$,
$X^{(l)}\equiv (X_p^{(l)},X_e^{(l)})$ with $X_a^{(l)}=\kappa_a
q_a^{(0)}+\sum_{l=1}^k \eta^{(l)}_a$ ($k$ runs from $1$ to
$n$), and $X^{(n+1)}\equiv (\kappa_p q_p^{(n+1)},\kappa_e
q_e^{(n+1)})=X^{(0)}$.

For $k=n+1$, we have 
\begin{equation}
\beta \frac{\partial}{\partial \beta}
\, {\hat P_e}  \rho^{(n+1)} =
-\beta \frac{\partial \Delta\beta \cdot U^c(X^{(n)})}{\partial
\beta} {\hat P_e} \rho^{(n+1)} + \beta {\hat P_e} 
{\tilde \rho}_{\beta}^{(n+1)}\,,
\label{rho-prime1}
\end{equation}
and, together with Eq.~(\ref{e-pimc}), we obtain for the energy
\begin{eqnarray}
\beta E &=& \frac{3}{2}(N_e+N_p) -
 \frac{1}{Z}
\frac{1}{\,\lambda_p^{3N_h}\lambda_e^{3N_e}} \int\limits_{V}
dq^{(0)} d\eta^{(1)} \dots d\eta^{(n)} \,
\sum_{\sigma}\sum_{P_e} (\pm
1)^{\kappa_{P_e}} \,{\cal S}(\sigma, {\hat P_e} \sigma')
\nonumber\\
&\times & \Bigg\{\sum_{k=1}^{n+1}\rho^{(1)}\dots\rho^{(l-1)}
\left[ \beta {\tilde \rho}_{\beta}^{(l)} -\beta \frac{\partial
\Delta\beta \cdot
U^c(X^{(l-1)})}{\partial \beta}\rho^{(l)}
 \right]
\rho^{(l+1)} \, \dots \,\rho^{(n)} {\hat P_e} 
\rho^{(n+1)} \Bigg\}\bigg|_{X^{(n+1)}=X^{(0)},\, \sigma'=\sigma}.
\label{pimc1}
\end{eqnarray}
This way the derivatives of the density matrix have been calculated and we turn to the next point:
Finding approximations for the high-temperature density matrices $\rho^{(l)}$.

\subsection{High-temperature asymptotics for the density matrix. Kelbg potential}
In this section we discuss approximations for the high-temperature density matrix which can be used for efficient direct PIMC simulations.
It involves effective quantum pair potentials $\Phi_{ab}$, which are approximated by the Kelbg potential. Here, we closely follow our
earlier work \cite{FiBoEbFo01} where details and further references can be found.

\subsubsection{Pair approximation and Kelbg potential}
The N-particle high-temperature density matrix is expressed in terms of two-particle
density matrices (higher order terms become negligible at sufficiently high
temperature, i.e. for large number $n$ of imaginary ``time slices'') given by
\begin{eqnarray}\label{rho_ab}
\rho_{ab}(q_{k,a},q'_{k,a}, q_{t,b}, q'_{t,b};\beta)
&=& \frac{(m_a m_b)^{3/2}}{(2 \pi \hbar \beta)^3}
\exp\left[-\frac{m_a}{2 \hbar^2 \beta} (q_{k,a} - q'_{k,a})^2\right]
\, \nonumber\\
&&\times \exp\left[-\frac{m_b}{2 \hbar^2 \beta} (q_{t,b} - q'_{t,b})^2\right]
\exp[-\beta \Phi^{OD}_{ab}]\,,
\end{eqnarray}
where $a,b=e,p$ and $k,t=1,...,N$ . 
This results from factorization into kinetic and interaction parts,
$\rho_{ab}\approx\rho_0^K\rho^{U^c}_{ab}$, which is exact in the classical
case, i.e. at sufficiently high temperature. The error made at finite 
temperature vanishes with the number of time slices as $n^{-2}$, cf. Ref.~\cite{FiBoEbFo01}.
The off-diagonal density matrix element (\ref{rho_ab}) involves an effective
pair interaction which is approximated by its diagonal elements according to
$\Phi^{OD}_{ab}(q_{k,a},q'_{k,a},q_{t,b}, q'_{t,b};\beta)\approx
\frac{1}{2}[\Phi_{ab}(q_{k,a}-q_{t,b}; \beta)+\Phi_{ab}(q'_{k,a}-q'_{t,b};\beta)]$,
for which we use the familiar Kelbg potential \cite{Ke63,kelbg}
\begin{eqnarray}
\Phi_{ab}(x_{ab};\beta) =
\frac{e_a e_b}{\lambda_{ab} x_{ab}} \,\left[1-e^{-x_{ab}^2} +
\sqrt{\pi} x_{ab} \left(1-{\rm erf}(x_{ab})\right) \right],
\label{kelbg-d}
\end{eqnarray}
where $x_{ab}=|q_{k,a}-q_{t,b}|/\lambda_{ab}$, and 
the error function is defined by ${\rm erf}(x)=\frac{2}{\sqrt{\pi}}\int_0^x
dt e^{-t^2}$. 
Note that the Kelbg potential is finite at zero distance which is a consequence of quantum
effects. The validity of this potential as well as of the diagonal 
approximation is restricted to temperatures substantially higher than
the binding energy \cite{afilinov-etal.04pre,ebeling_sccs05} which puts 
another lower bound on the number of time slices $n$. 
For completeness we also note other effective potentials, e.g. Refs.~\cite{KTR94, deutsch_77}, as well as recently derived improved 
versions that are applicable to strong coupling \cite{afilinov-etal.04pre, ebeling_sccs05}.

Summarizing, we can conclude that, with the approximations (\ref{rho_ab},\ref{kelbg-d}), each of the 
high-temperature density matrix factors on the r.h.s. of Eq. (\ref{rho-pimc}), carries an error of the order $1/(n+1)^2$,
\begin{eqnarray}
\rho^{(l)}=\rho_0^{(l)}e^{-\Delta \beta
U(X^{(l-1)})}\delta(X^{(l-1)}-X^{(l)})+{\cal O}[(n+1)^{-2}],
\label{kel}
\end{eqnarray}
where $\rho_0^{(l)}$ is the kinetic density matrix, and $U$ denotes
the sum of all interaction energies, each consisting of the
respective sum of pair interactions given by Kelbg potentials,
$U(X^{(l)})=U_{pp}(X_p^{(l)})+U_{ee}(X_e^{(l)})+U_{ep}(X_p^{(l)},X_e^{(l)})$.

\subsubsection{Estimator for the total energy}
Let us now return to the computation of thermodynamic functions and derive the
final expressions, following from Eq.~(\ref{kel}), that will be used in the simulations. First, we note that in Eq.~(\ref{pimc1}), special care has to be taken in performing the derivatives of the Coulomb potentials with respect to $\beta$: Products $\beta\frac{\partial \Delta\beta \cdot U^c(X^{(l-1)})}{\partial \beta}$
have a singularity at zero inter-particle distance which is integrable but leads to difficulties in the simulations.
To assure efficient simulations we transform the e-e, p-p and e-p contributions in the following way:
\begin{eqnarray}
&&\hspace*{-2cm}
\left\langle X^{(l-1)}\left|e^{-\Delta \beta {\hat K}} \right|X^{(l)}\right\rangle 
\left[-\beta\frac{\partial}{\partial\beta}\left(\Delta \beta U^c(X^{(l-1)})\right) \right]
\nonumber\\[2ex]
&&\approx  \int_0^1  d\alpha \int d\tilde{X}^{(l-1)} \left\langle
X^{(l-1)}\left|e^{-\Delta \beta \alpha{\hat K}}\right|\tilde{X}^{(l-1)}\right\rangle
\left[-\beta\frac{\partial}{\partial\beta} \left(\Delta \beta
U^c(\tilde{X}^{(l-1)})\right) \right]
\nonumber\\[2ex]
 &&\qquad \qquad \qquad\times \left\langle \tilde{X}^{(l-1)}\left|e^{-\Delta \beta
(1-\alpha){\hat K}} \right|X^{k}\right\rangle +{\cal O}\left[(n+1)^{-2}\right]
\, \nonumber\\[2ex]
&&\approx  \left\langle X^{(l-1)}\left|e^{-\Delta \beta {\hat K}}
\right|X^{(l)}\right\rangle \left[-\beta\frac{\partial}{\partial\beta}
\left(\Delta \beta U(X^{(l-1)})\right) \right] + {\cal
O}\left[(n+1)^{-2}\right]. \label{uep}
\end{eqnarray}
where $\langle \dots | \dots \rangle$ denotes the scalar product. This means, within the standard error of our approximation, ${\cal
O}\left(n^{-2}\right)$, we have replaced the sum of the
Coulomb potentials $U^c$ by the corresponding sum of Kelbg
potentials $U$, which is much better suited for MC simulations.
This result coincides with expressions, which can be obtained
if we first choose an approximation for the high-temperature
density matrices $\rho^{(l)}$ using the Kelbg potential and then take the derivatives.

Thus, our final result for the energy is
\begin{eqnarray}
\beta E &=& \frac{3}{2}(N_e+N_p) + \frac{1}{Z}\frac{1}{\,\lambda_p^{3N_p} \lambda_e^{3N_e}}
\sum_{s=0}^{N_e}  \int\limits_{V} dq^{(0)}
d\eta^{(1)} \dots d\eta^{(n)} \,\rho_{s}(q^{(0)}, \eta^{(1)} \dots \eta^{(n)}, \beta)
\nonumber\\[2ex]
&&\times \Bigg\{ \sum_{l=0}^{n}\Bigg[
\sum_{k=1}^{N_p}\sum_{t=1}^{N_e} \Phi_{ep}(|x^l_{kt}|) +
\sum_{k<t}^{N_p} \Phi_{ee}(|r^l_{kt}|)
+\sum_{k<t}^{N_e} \Phi_{pp}(|q^l_{kt}|)
 \Bigg]
\nonumber\\[2ex]
&& \quad + \sum_{l=1}^{n}\Bigg[
 \sum_{k=1}^{N_p}\sum_{t=1}^{N_e}
D(x^l_{kt}) \frac{\partial \Delta\beta\Phi_{ep}}{\partial
|x^l_{kt}|}
 + \sum_{k<t}^{N_e}
C(r^l_{kt}) \frac{\partial \Delta\beta\Phi_{ee}}{\partial
|r^l_{kt}|} + \sum_{k<t}^{N_e} C(q^l_{kt}) \frac{\partial
\Delta\beta\Phi_{pp}}{\partial |q^l_{kt}|}
 \Bigg]
\nonumber\\[2ex]
&& \quad - \frac{1}{{\rm det} ||\psi^{n,0}_{kt}||_{s}}
\frac{\partial{\rm \,det} || \psi^{n,0}_{kt} ||_{s}}{\partial
\beta} \Bigg\},
\label{energy}
\end{eqnarray}
with the definitions
\begin{eqnarray}
C(r^l_{kt}) &=& \frac{\langle r^l_{kt}|y^l_{kt}\rangle}{2|r^l_{kt}|}, 
\qquad C(q^l_{kt}) = \frac{\langle q^l_{kt}|\tilde{y}^l_{kt}\rangle}{2|q^l_{kt}|},
\nonumber\\[2ex]
D(x^l_{kt}) &=& \frac{\langle
x^l_{kt}|y^l_{p}-\tilde{y}^l_{t}\rangle}{2|x^l_{kt}|},
\qquad
\Psi_{ab}(x) \equiv  \Delta\beta \frac{\partial [\beta'\Phi_{ab}(x,\beta')]}{\partial\beta'|_{\beta'=\Delta\beta}}. 
\nonumber 
\end{eqnarray}
Here we introduced the following notation for the differences of two coordinate vectors
\begin{eqnarray}
q_{kt} &\equiv & q_{k,p}-q_{t,p}, \qquad r_{kt}\equiv q_{k,e}-q_{t,e}, \qquad x_{kt}\equiv q_{k,e}-q_{t,p},
\nonumber\\
r^l_{kt} &=& r_{kt}+y_{kt}^l, \qquad q^l_{kt}=q_{kt}+\tilde{y}_{kt}^l, \qquad x^l_{kt}\equiv
x_{kt}+y^l_k-\tilde{y}^l_t,  
\nonumber\\
y^l_{kt} &\equiv &y^l_{k}-y^l_{t},
\qquad \tilde{y}^l_{kt}\equiv \tilde{y}^l_{p}-\tilde{y}^l_{t},
\nonumber
\end{eqnarray}
with $y^l_t=\Delta\lambda_e\sum_{k'=1}^{l}\eta^{(k')}_t$ and  $\tilde{y}^l_k=\Delta\lambda_p\sum_{k'=1}^{l}\tilde{\eta}^{(k')}_k$.
The density matrices $\rho_{s}$ appearing in Eq.~(\ref{energy}) are given by
\begin{eqnarray}
\rho_{s} = C^s_{N_e} \, e^{-\beta U}
{\rm det}||\psi^{n,0}_{kt}||_{s} 
\prod\limits_{l=1}^n \prod\limits_{k=1}^{N_e}
\prod\limits_{t=1}^{N_p} \phi^l_{k} \,\tilde{\phi}^l_{t}  ,
\label{rho_s} \, 
\end{eqnarray}
where $U$ is the total interaction energy comprised of contributions from all time slices,
\begin{equation}
U = \frac{1}{n+1}\sum_{l=0}^{n} \left\{ U_e(X_e^{(l)}, \Delta \beta)
+U_p(X_p^{(l)},\Delta\beta)+
U_{ep}(X_p^{(l)},X_e^{(l)},\Delta\beta) \right\} 
\label{Uenr} \, 
\end{equation}
and we defined
\begin{eqnarray}
 \phi^l_{t} &\equiv &\exp[-\pi |\eta^{(l)}_t|^2],  \qquad
\tilde{\phi}^l_{k}  \equiv \exp[-\pi |\tilde{\eta}^{(l)}_k|^2],
\nonumber\\[2ex]
||\psi^{n,0}_{kt}||_{s} &=& \left \|e^{-\frac{\pi}{\Delta\lambda_e^2}
\left|(r_k-r_t)+ y_k^n\right|^2}\right\|_s \times \left\|e^{-\frac{\pi}{\Delta\lambda_e^2} \left|(r_k-r_t)+
y_k^n\right|^2} \right\|_{N_e-s}.
 \label{psi}
\end{eqnarray}
Notice that the density matrix (\ref{rho_s}) does not contain an explicit sum over the permutations and, thus, no sum of terms
with alternating sign. Instead, the whole exchange problem is contained in the determinant (\ref{psi}) which is a product of exchange matrices of electrons 
where $s$ 
denotes the number of electrons 
having the same spin projections (for more details, we refer to Ref.~\cite{filinov76}). This grouping of terms with different signs into the spin determinant is similar to ``blocking'' algorithms, e.g. \cite{egger_blocking}, and allows to substantially weaken the Fermion sign problem.

\subsection{Path integral Monte Carlo procedure for jellium}
The above formulas have been applied successfullly to multi-component dense quantum plasma simulations \cite{FiBoEbFo01,Bonitz_PRL05,Filinovsccs,levashov_jpa_06},
  to the electron-hole plasmas in semiconductors \cite{filinov_jpa_03,eh} as well as to the quark-gluon plasma \cite{filinov_cpp12,filinov_prc13}.
It is, therefore, desirable to retain the same program also for the simulation of jellium where the positive component (protons)  
 is treated as a homogeneous static background. This has the advantage that both components are treated consistently, in particular, the background 
contribution will be automatically adjusted to the chosen particle number (which would not be the case if we would use the known corrections for 
the case of an infinite system).

To perform the transition to the case of jellium with minimal changes we put the potential energy contributions of the protons---the p-p and p-e interactions in the exponent of the high-temperature density matrices---to zero. 
Thus expression (\ref{Uenr}) is reduced to 
\begin{eqnarray}
U=\frac{1}{n+1}\sum_{l=0}^{n} U_e\left(X_e^{(l)}, \Delta \beta \right).
\label{JU}
\end{eqnarray}
At the same time, the interaction terms $\Psi_{ep}$ and $\Psi_{pp}$ are retained as they produce the energy contribution of the ``background''. With this, the proton component is treated as an ideal gas of given density and temperature with the proton number always matching that of the electrons, guaranteeing charge neutrality.
With these trivial changes, expressions~(\ref{energy}), (\ref{rho_s}) and (\ref{JU}) are well suited for
efficient fermionic PIMC simulations of jellium using standard Metropolos Monte Carlo techniques (see,
 e.g.,~\cite{zamalin,binder96,numbook}). In our Monte Carlo scheme we use three different types of moves, where either electronic $q_{t,e}$ or positive charge 
coordinates $q_{p,h}$ or the individual electronic beads, $\eta_t^{(k)}$,
are moved until convergence of the calculated values is reached. 

Computer simulations of disordered systems, such as plasmas, require an accurate account of long-range Coulomb
forces which strongly affect the thermodynamic and transport properties. 
Accurate computer simulations require sometimes up to a million particles in
the main Monte Carlo computation cell. Moreover, the larger the number of charged 
particles in the main cell, the more acute is the problem of an efficient evaluation of the 
Coulomb contribution. To reduce the effects of finite number of particles 
usually periodic boundary conditions (PBC) are imposed on the main Monte Carlo cell. 
An essential problem in computer simulations of such systems is to properly combine the accurate account of long-range Coulomb forces with PBC. 
One way to do this is the Ewald summation method.
However, the usual Ewald procedure, invokes an artificial non-isotropic electric ``crystalline field'' in the spatially uniform and isotropic
 Coulomb system. Moreover periodicity artifacts are heavy processor load imposed by Ewald 
summation procedure in computer simulations. Recently a modified Ewald scheme has been derived that avoids the ``crystalline field'' by suitable 
angle averages and was applied to Coulomb systems~\cite{Yakub}. 
Here we use this modified Ewald scheme for particles interacting via Kelbg potential. We expect that the same procedure is justified because the Kelbg 
and Coulomb potentials have identical long range asymptotics. 

To realize this concept we identically rewrite the Kelbg potential as the sum of a short range part, 
$\Delta\Phi_{ab}(x_{ab};\beta)= \Phi_{ab}(x_{ab};\beta) - \frac{e_a e_b}{\lambda_{ab} x_{ab}} $, 
and the long-range Coulomb potential 
\begin{eqnarray}
\Phi_{ab}(x_{ab};\beta)=\Delta \Phi_{ab}(x_{ab};\beta)+
\frac{e_a e_b}{\lambda_{ab} x_{ab}},
\label{kelbg-delta}
\end{eqnarray} 
Following Reference \cite{Yakub}, after averaging over all orientations of the main Monte Carlo cell, 
we obtain a potential $\tilde{\Phi}_{ab}(x_{ab};\beta)$ that accounts for PBC, 
for distances $x_{ab}<x_m$,
\begin{eqnarray}
\tilde{\Phi}_{ab}(x_{ab};\beta) &=& \Delta \Phi_{ab}(x_{ab};\beta)+\theta(x_m-x_{ab})
\frac{e_a e_b}{\lambda_{ab} x_{ab}}\,\left[1 + \frac{1}{2}\left(\frac{x_{ab}}{x_m}\right)
\left[\left(\frac{x_{ab}}{x_m}\right)^2-3 \right]\right] 
\nonumber\\ 
&=& \Phi_{ab}(x_{ab};\beta)+ \left(\theta(x_m-x_{ab})-1\right)
\frac{e_a e_b}{\lambda_{ab} x_{ab}}\,+ \frac{1}{2}\theta(x_m-x_{ab})\frac{e_a e_b}{\lambda_{ab} x_m}\,
\left[\left(\frac{x_{ab}}{x_m}\right)^2-3 \right],
\label{kelbg-pbc}
\end{eqnarray} 
where $\theta(x_m-x_{ab})$ is the Haviside step function 
The parameter $x_m$ defined by $\frac{4}{3}\pi x_m^3 =L^3$, is the radius of the volume-equivalent 
sphere of the main Monte Carlo cell of volume $L^3$. 

The effective pair potential $\tilde{\Phi}$, Eq.~(\ref{kelbg-pbc}), has the following properties:
\begin{itemize}
 \item At small distances, $x < x_m$, it tends to the Kelbg potential with PBC corrections arising 
from its long-range Coulomb asymptotic.
 \item At large distances, $x > x_m$, due to the the coincidence of the long-range asymptotics of the 
Kelbg and Coulomb potentials, the effective potential $\tilde{\Phi}$ tends zero. 
\end{itemize}

In expressions~(\ref{energy}) and (\ref{JU}) and related calculations 
we now replace the potential $\Phi_{ab}(x_{ab};\beta)$ by $\tilde{\Phi}_{ab}(x_{ab};\beta)$, thereby 
accounting for PBC effects. Let us note that, to the first term in curly brackets in 
Eq.~(\ref{energy}), we have to add a constant equal to 
$-\frac{3 e_a e_b (N_e+N_p)}{16 \pi \lambda_{ab} x_m}$ \cite{Yakub}. An analogous constant 
shift for the sum Eq.~(\ref{JU}) is not important and is omitted. 

The main contribution to the path integral representation of the partition function comes from configurations for
which the typical size of the clouds of electronic  beads is of the order of the thermal wave length $\lambda$ of the electrons. 
In the simulations bewlow we use up to about one hundred electrons 
in the basic MC cell. Due to this limitation we have a related restriction on the size of the MC cell, for a given density. 

Let us note another important improvement.
In our previous calculations determinants of the exchange matrices were only computed for particles belonging to the basic Monte
Carlo cell. However, in the case of a high degeneracy, $n\lambda^3 \gg 1$, the thermal wave length (and the typical size of the electronic clouds of beads) may easily exceed the  size $L$ of the MC cell. So beads of electrons belonging to the basic MC cell can penetrate into neighboring images of the main cell,
and, vice versa, electronic beads from neighboring cells can extend into the basic cell. 
This requires a modified treatment of exchange in the PIMC simulations: it is necessary to include exchange effects between particles in the main MC cell and their
images in the neighboring cells as well. Therefore, in the present calculations we take into account the exchange interactions of
electrons with the electrons from the  ($3^3-1$) nearest-neighbor cells. (As a second step, this could be extended to the $(5^3-1)$
nearest and next nearest-neighbor cells, until convergence is reached. Due to the rapidly increasing computational effort this has not been 
done in the present work and will be studied in a forthcoming work.)
This is the main modification to the PIMC algorithm compared to our earlier work.

For the present simulations  we varied, both, the particle number and the number of beads in the range of $N_e=N_p=50 \dots 100$ particles and $n=20\dots 90$ beads. 
This interval of $N$ and $n$ is used to perform a finite size scaling to the macroscopic limit, i.e. an approximate extrapolation to $n \to \infty$ and $N\to \infty$.
Furthermore, as in previous PIMC studies of the finite temperature UEG we study the polarized and unpolarized cases.
This substantially simplifies the sums over the total electron spin $s$. 
%
Our simulations with the improved treatment of the electronic exchange were first tested for an ideal plasma. The agreement with the known analytical results for an ideal finite temperature Fermi gas is found to be very good, up to densities where the degeneracy parameter $n\lambda^3$ reaches values of the order of $100$. 
This is demonstrated below in Figs.~\ref{fig:Pol}, \ref{fig:UnPol}. 
%
\section{Simulation results}\label{simulations}
We now apply the theoretical scheme developed in the preceding
sections to unpolarized and polarized 
jellium. The thermodynamic state of the electron gas is characterized by the dimensionless density and temperature.
Below, the density of the electrons is characterized by the Brueckner parameter, $r_s=a/a_B$, defined as the ratio of the mean distance
between particles, $a=\left[\frac{3}{4\pi n_e}\right]^{1/3}$, and the Bohr radius, $a_B$, where $n_e$ is electron density. Temperature will 
be given in units of the Fermi energy, $\Theta=k_BT/E_F$.
\begin{figure}[htpm]
\includegraphics[width=8.1cm,clip=true]{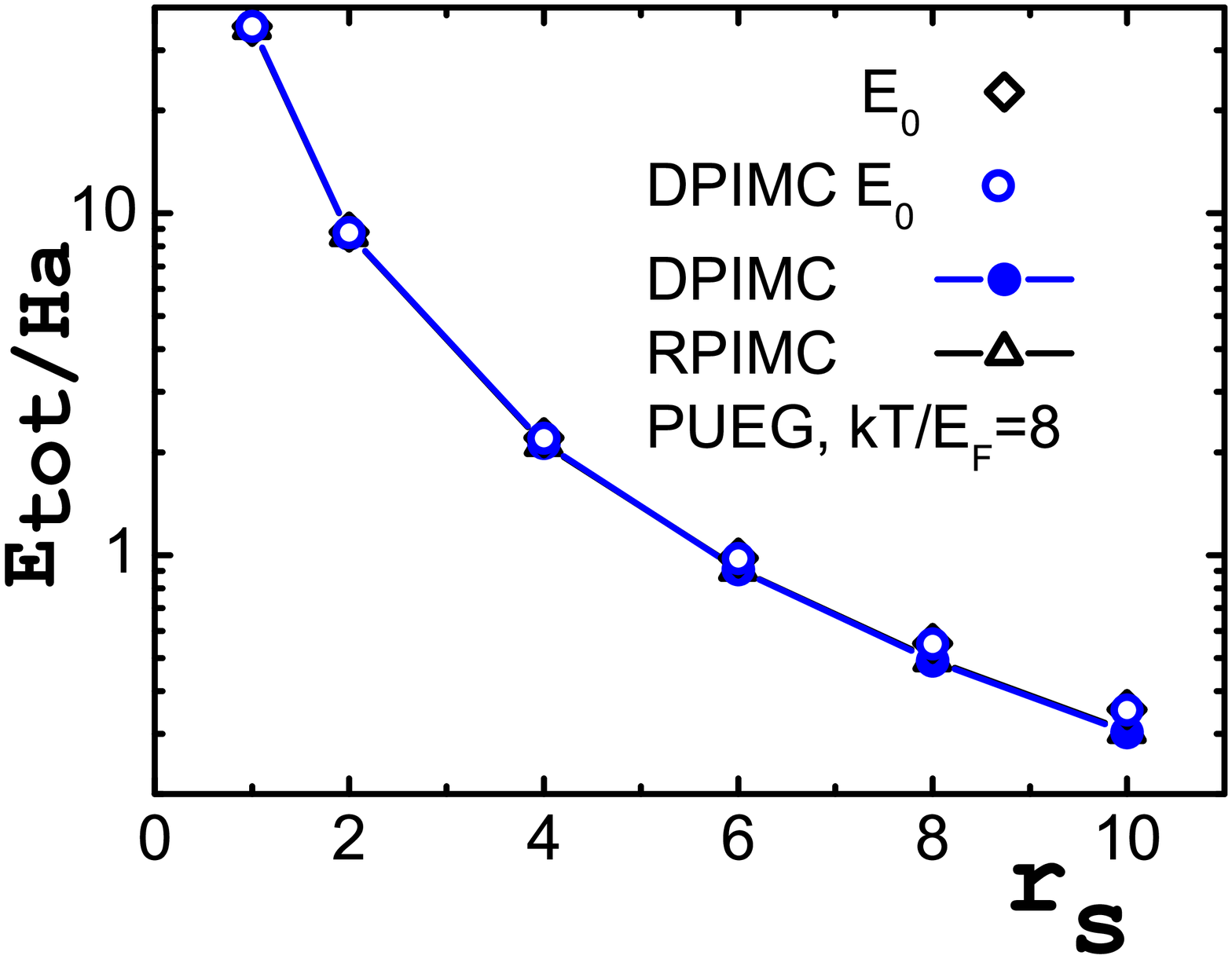}
\includegraphics[width=8.1cm,clip=true]{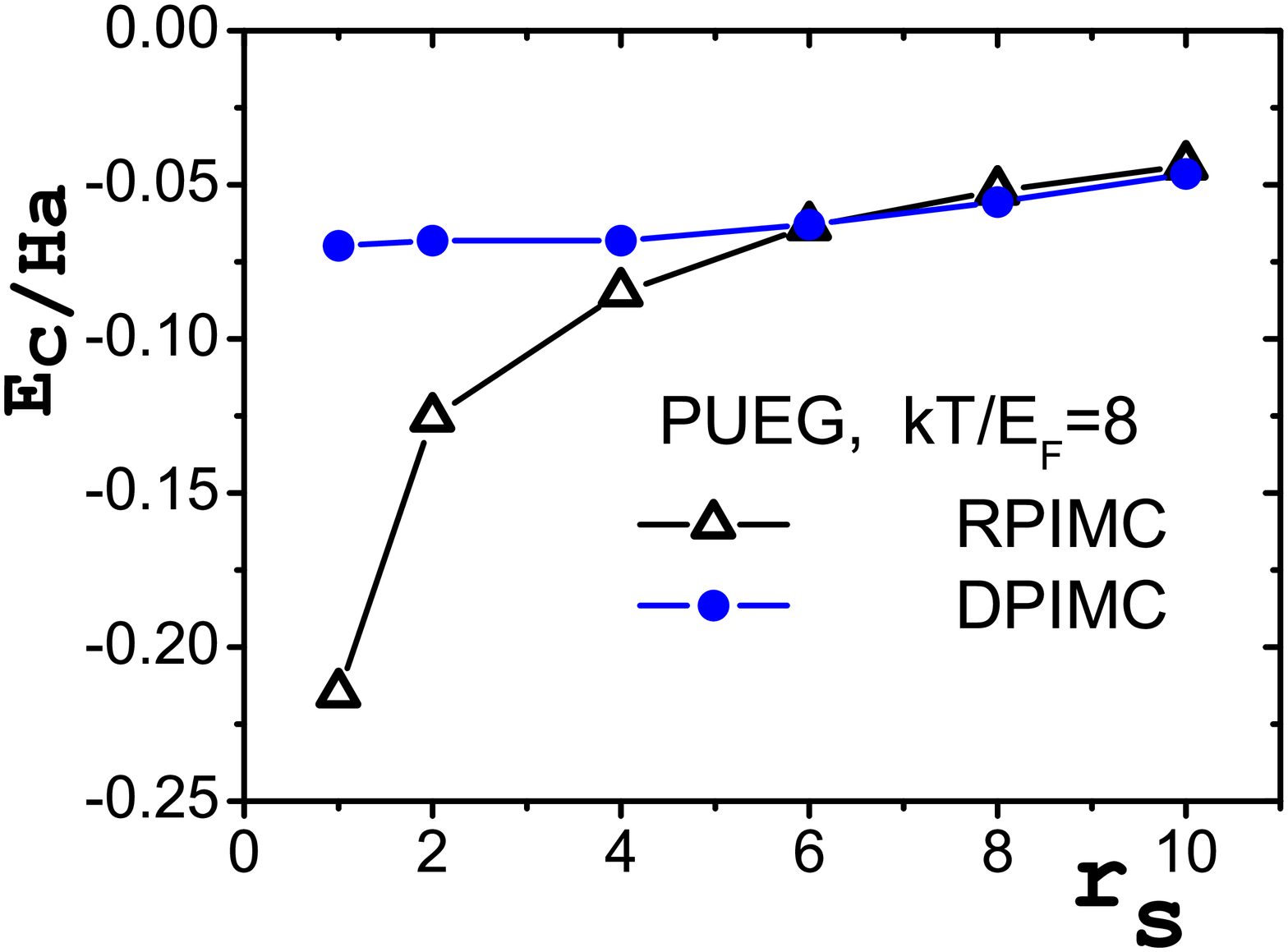} 
\\\includegraphics[width=8.1cm,clip=true]{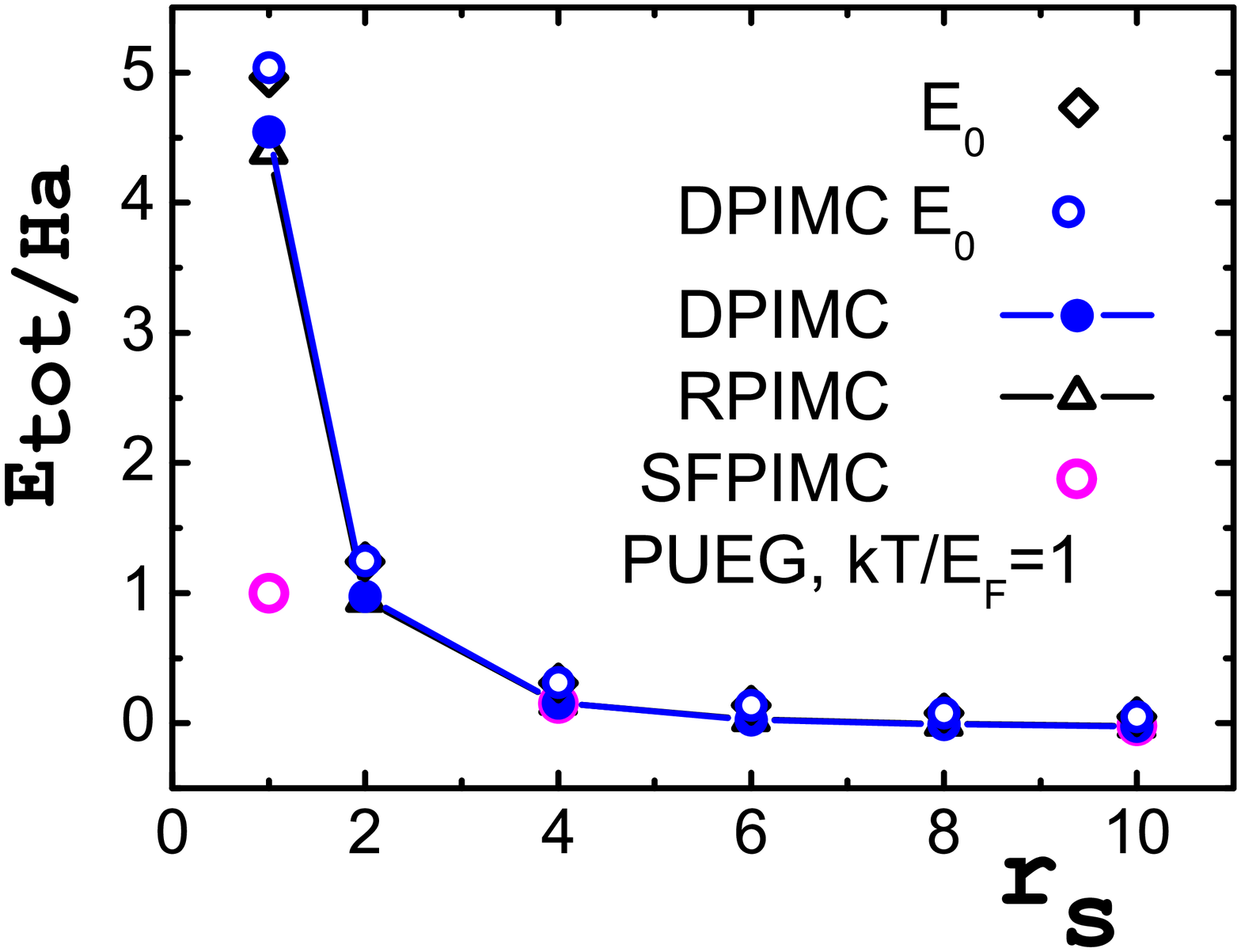}
\includegraphics[width=8.1cm,clip=true]{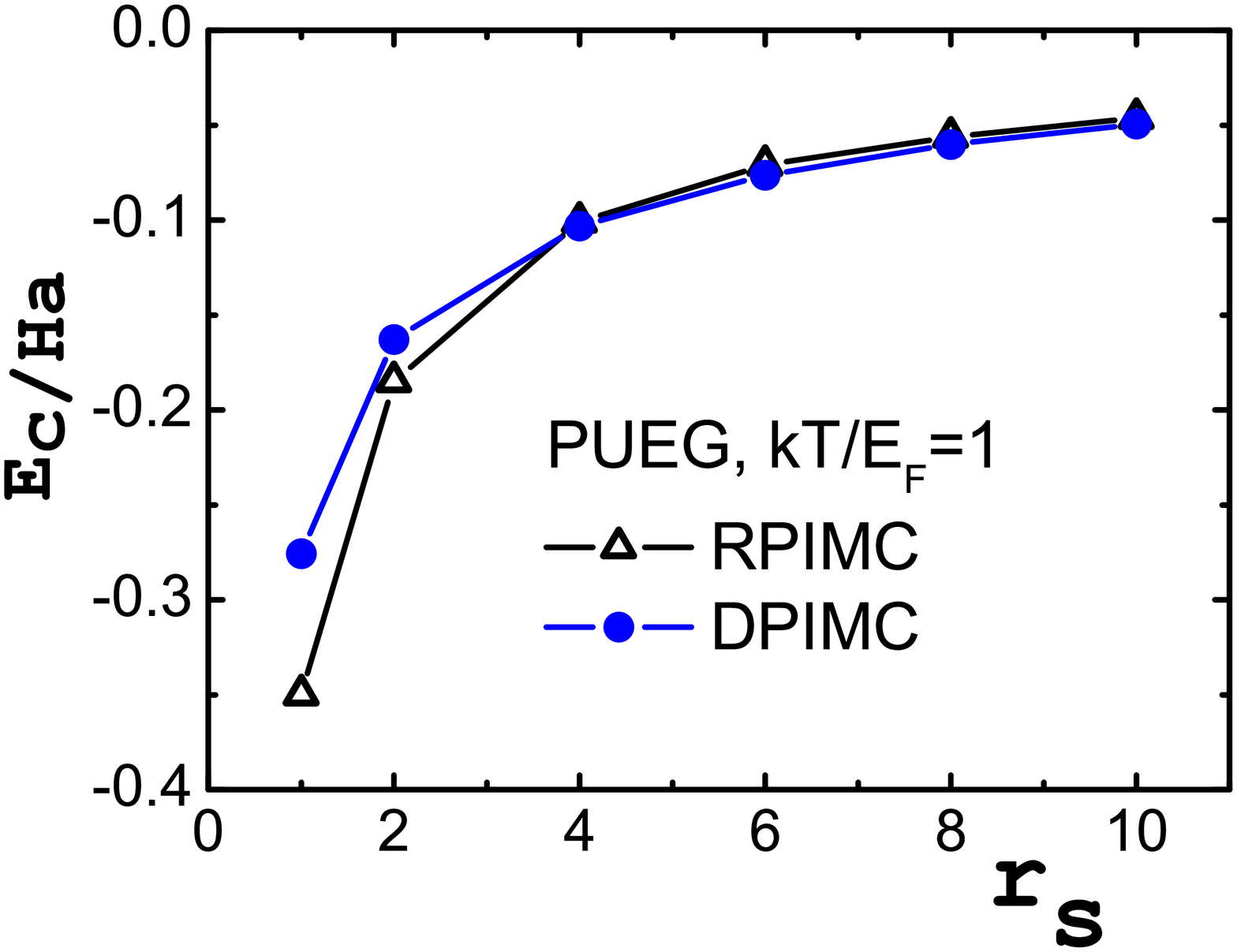} 
\includegraphics[width=8.1cm,clip=true]{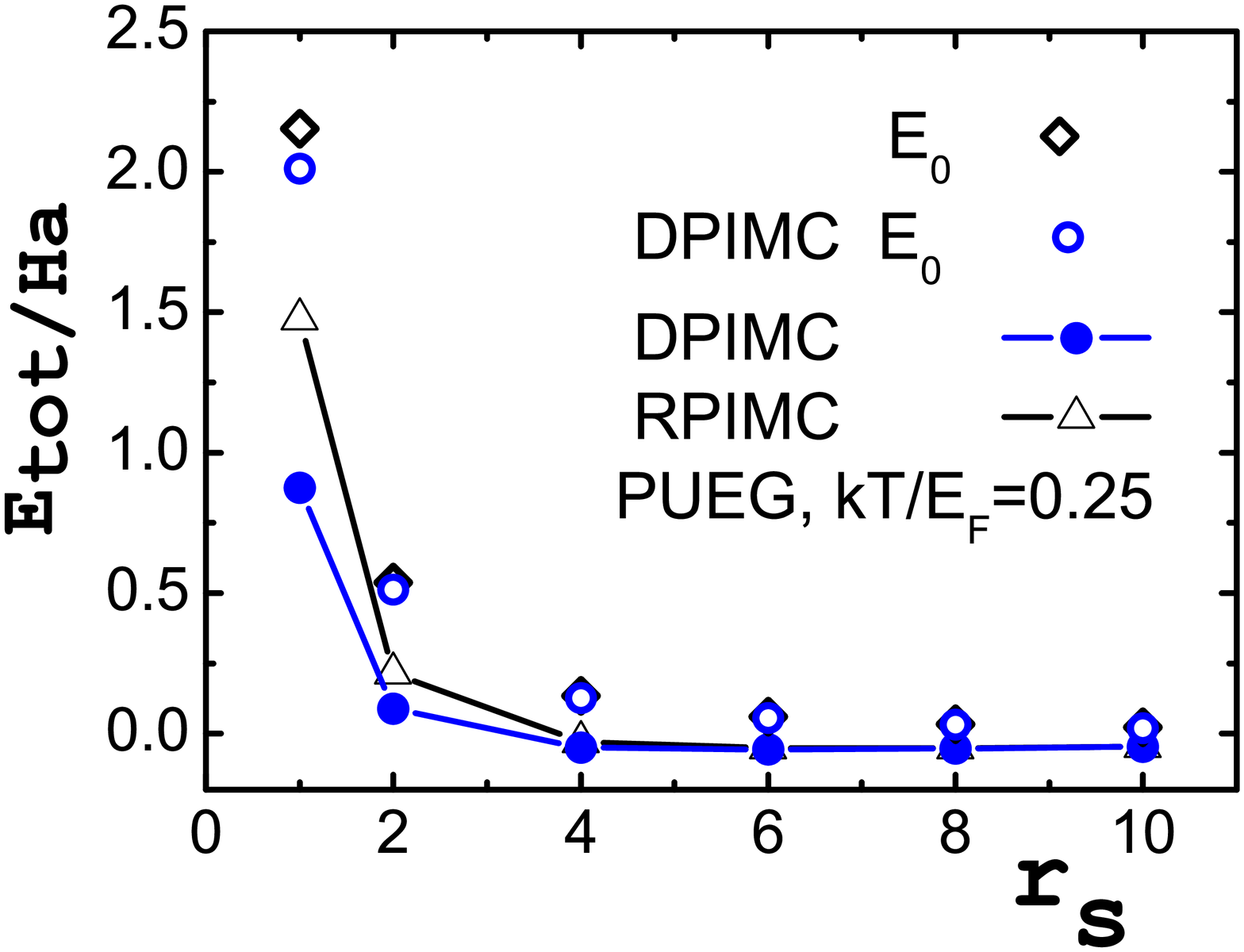}
\includegraphics[width=8.1cm,clip=true]{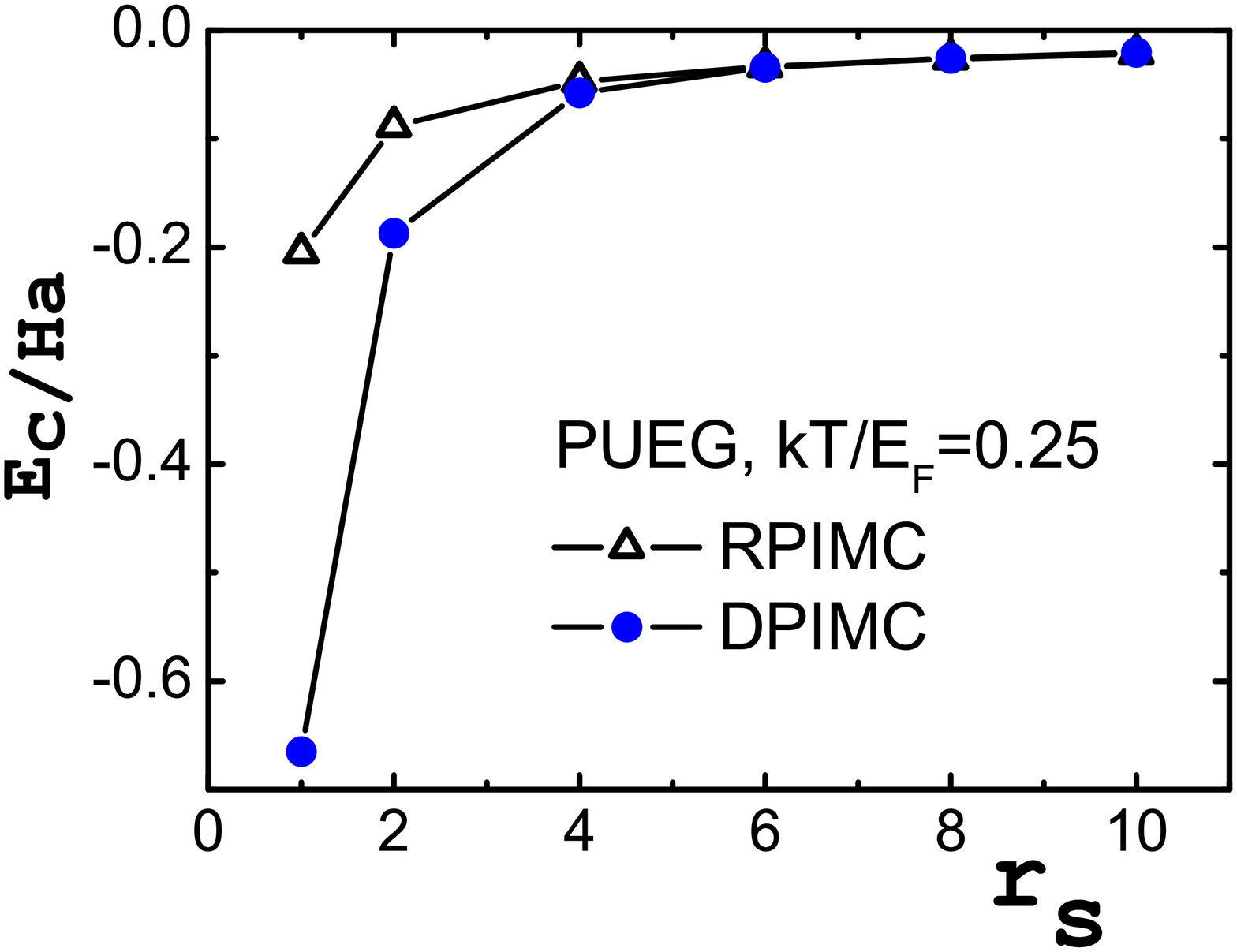} 
\includegraphics[width=8.1cm,clip=true]{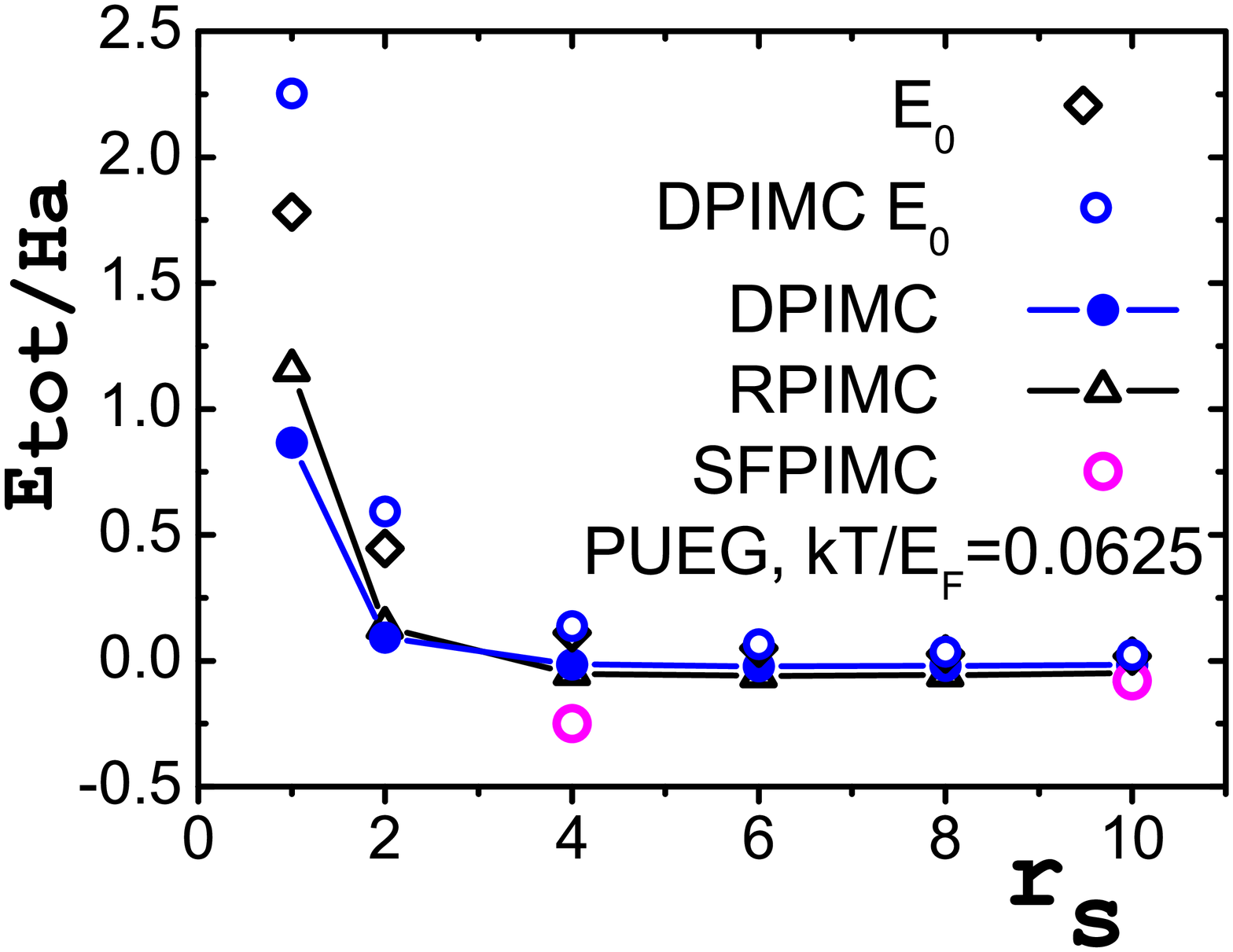}
\includegraphics[width=8.1cm,clip=true]{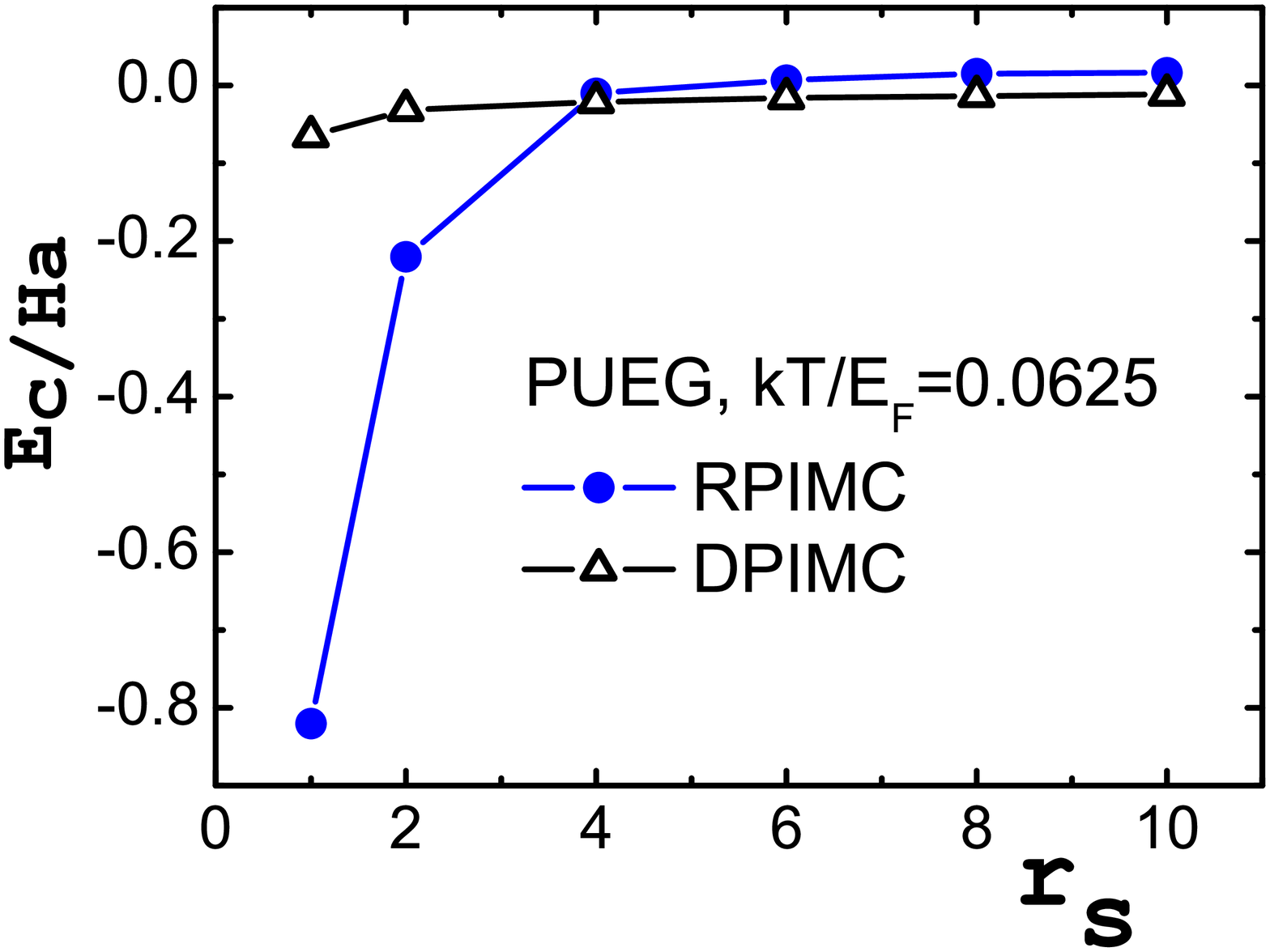}
\caption{(Color online) 
Total energy per particle (left column) and correlation energy per particle (right column) for a polarized ideal ($E_0$) and interacting  electron gas (PUEG) with temperatures ranging from $\Theta=8$ to $\Theta=0.0625$, 
see text in the graphs. Comparison of restricted PIMC (RPIMC, Ref.~\cite{brown13}), fermionic PIMC 
(SFPIMC, open circles \cite{brown13}) and the present DPIMC results. 
The correlation energy is given by $E_c = E_{tot}-E_0-E_{x,HF}$ using the Hartree-Fock energy  of Ref. \cite{brown13}, see main text.)
\label{fig:Pol}
}
\end{figure}
%
%
We now present our fermionic path integral Monte Carlo simulation results for the energy 
 of the uniform electron gas, based on Eq.~(\ref{energy}), with the simplifications 
discussed above. We use the same density and temperature interval that was studied in Ref.~\cite{brown13}: 
$r_s \ge 1$ and $\Theta = 0.0625 \dots 8$. Higher densities are not considered since there the sign problem is too severe and the data are not reliable. 
Figures~\ref{fig:Pol} and ~\ref{fig:UnPol} show 
the results for a polarized electron gas and for the unpolarized case, respectively. 
The left columns  show the total energy for four different temperatures.
In the right columns of Figs.~\ref{fig:Pol} and ~\ref{fig:UnPol} we present the results for the correlation energy. 
 It was calculated as $E_c=E_{tot}-E_0-E_{x,HF}$ subtracting from the total energy the kinetic energy and the finite temperature Hartree-Fock (mean field plus exchange) energy of Ref.~\cite{brown13}. Since  $E_{tot}$ and $E_0+E_{x,HF}$ are of the same order of magnitude,
the reminder (the correlation energy) is very susceptible to different approximations and allows for sensitive comparisons of our results to those of Ref.~\cite{brown13}. 

Consider first the polarized case, Fig.~\ref{fig:Pol}. The much improved treatment of fermionic exchange (see above) is confirmed by the four plots for the total energy (left columns ). The curve denoted ``$E_0$''  shows the (kinetic) energy of an ideal polarized Fermi gas at finite temperature computed from the relevant Fermi integral. The curves ``DPIMC $E_0$'' show the same result obtained with our fermionic 
PIMC simulations, i.e. in these simulations all interaction terms were turned off. For the two highest temperatures, $\Theta = 8$ and $\Theta=1$, the 
results coincide practically for all densities. For $\Theta=0.25$ small deviations are seen at the highest density, $r_s=1$, whereas for $\Theta=0.0625$ 
deviations are visible up to $r_s \sim 3$. This behavior is a very good test for the order of magnitude of the error and for the reliability of our simulations.

If interactions are included, the total energy is lower compared to the ideal case. The curves are denoted by ``DPIMC'' and are compared to the associated 
restricted PIMC data of Brown {\em et al.} \cite{brown13} labeled ``RPIMC''. Again, the agreement over the entire temperature interval is rather good. 
An exception is the point $r_s=1$, with the highest density. For $\Theta=0.25$ our data are substantially lower than RPIMC. Interestingly, for $\Theta=0.625$ the 
agreement is significantly better although the accuracy of our results is most likely lower due to the increased sign problem. 
 Also, notice the fermionic (``signful'') PIMC simulations of Ref. \cite{brown13} labeled ``SFPIMC'' that are plotted with the open (pink) circles. For low densities we observe good agreement with our data, whereas for $r_s \lesssim 4$ there are significant deviations. At these points the SFPIMC data already carry large error bars and we expect that our results, due to the various improvements of the simulation approach, are more accurate.

In contrast to the rather good agreement of the toal energies there is a quit large deviation of our correlation energy from the data of Ref.~\cite{brown13}. For different densities the deviations can be positive or negative: for hight temperatures ($\Theta \ge 1$) our values are generally higher, whereas for $\Theta = 0.25, 0.0625$ our results are substantially lower at high densities $r_s \le 4$.  
The origin of these substantial deviations are still under investigation. We believe that they are, at least, partially due to finite size correction which are applied to the data Ref.~\cite{brown13} but not to our results.

Consider now the energies in the case of an unpolarized electron gas, see Fig.~\ref{fig:UnPol}. Again, the ideal system consitute a benchmark for the treatment of fermionic exchange in our simulations. 
Here the agreement with the analytical results is slightly worse than in the polarized case before, due to poorer convergence. Deviations are seen already for $\Theta=1$ and $r_s=1$ and continue to grow for $\Theta=0.25$ and $\Theta=0.0625$. The comparison of the results for the interacting electron gas with the RPIMC data of Ref.~\cite{brown13} is similar as in the polarized case. At temperatures $\Theta \gtrsim 1$ the agreement is very good. Deviations start to appear for $\Theta=1$, at $r_s=1$. At the lowest temperature, $\Theta=0.0625$, deviations are observed for all densities. Comparing again with the fermionic (SFPIMC) data of reference \cite{brown13}, cf. the data point at $\Theta=r_s=1$ we observe dramatic deviations because the SFPIMC results carried a very large error. Interestingly, in our case the error appears to be much smaller which we attribute to the improved treatment of fermionic exchange (using the nearest neighbor cells) and the use of the exchange determinant which constitutes an efficient approach to group different exchange contributions together which partially compensate each other and increase the sign in the simulations. The difference in the correlation energies of our results and compared to Ref. \cite{brown13} are similar as in the polarized case.

\begin{figure}[htpm]
\includegraphics[width=8.1cm,clip=true]{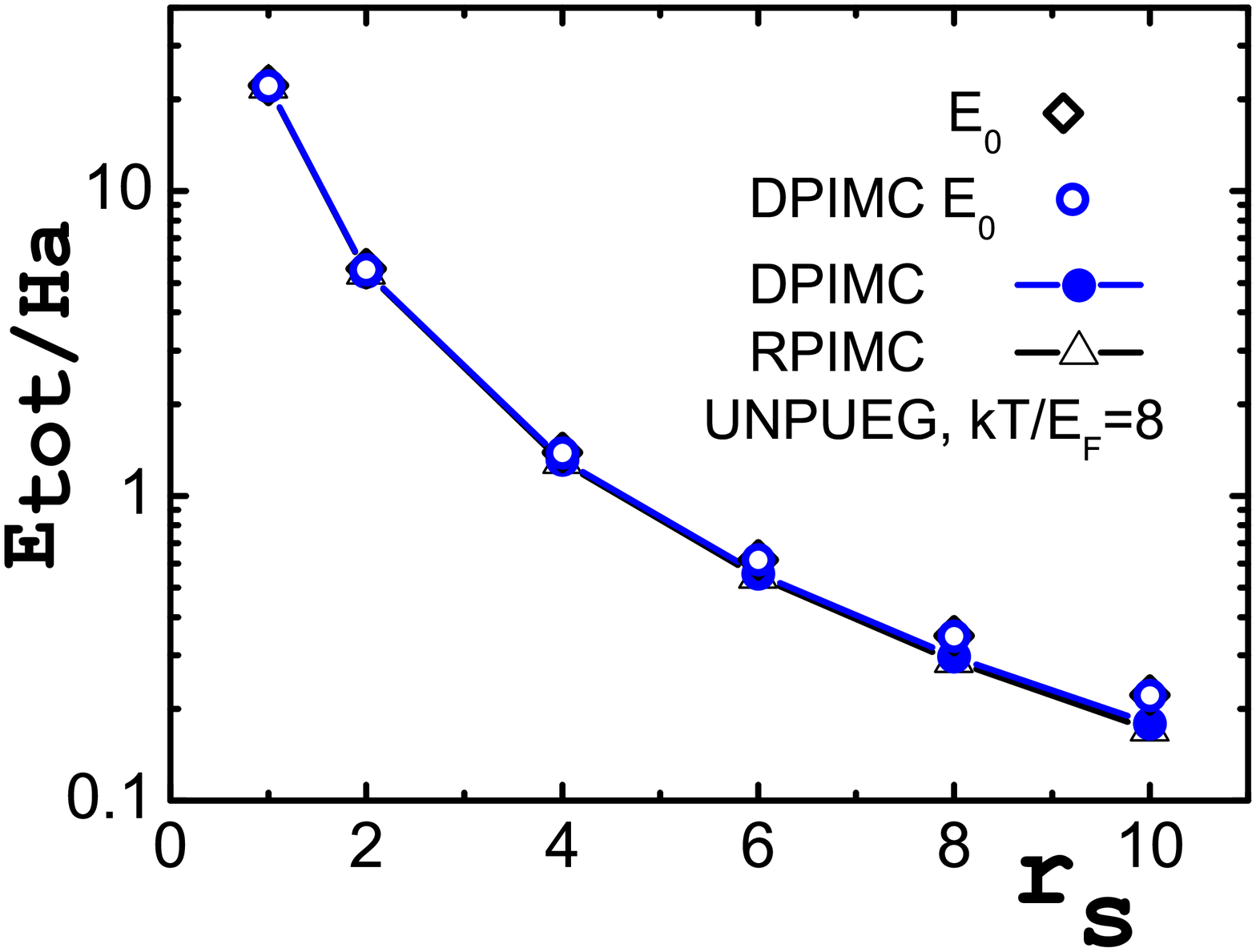}
\includegraphics[width=8.1cm,clip=true]{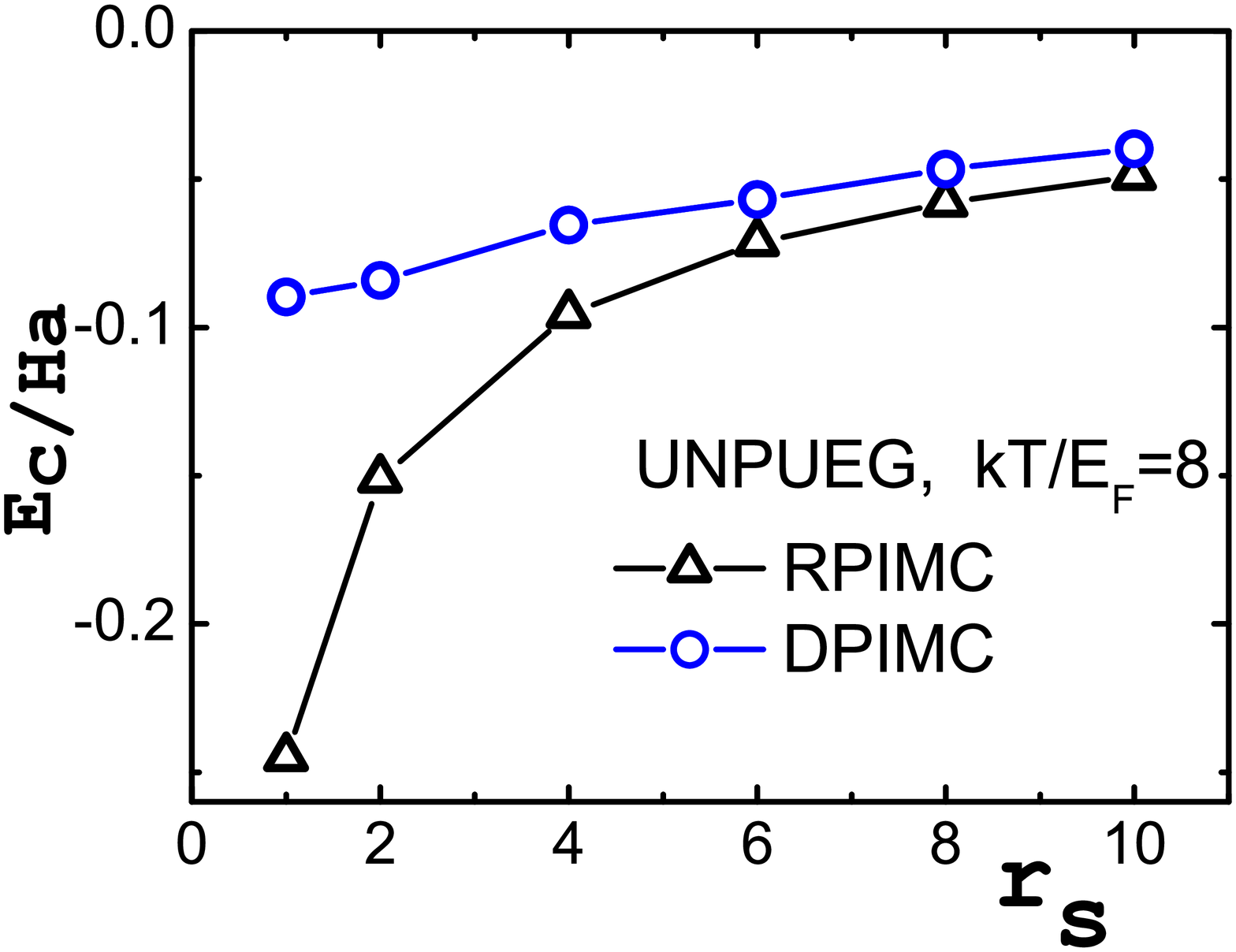}
\includegraphics[width=8.1cm,clip=true]{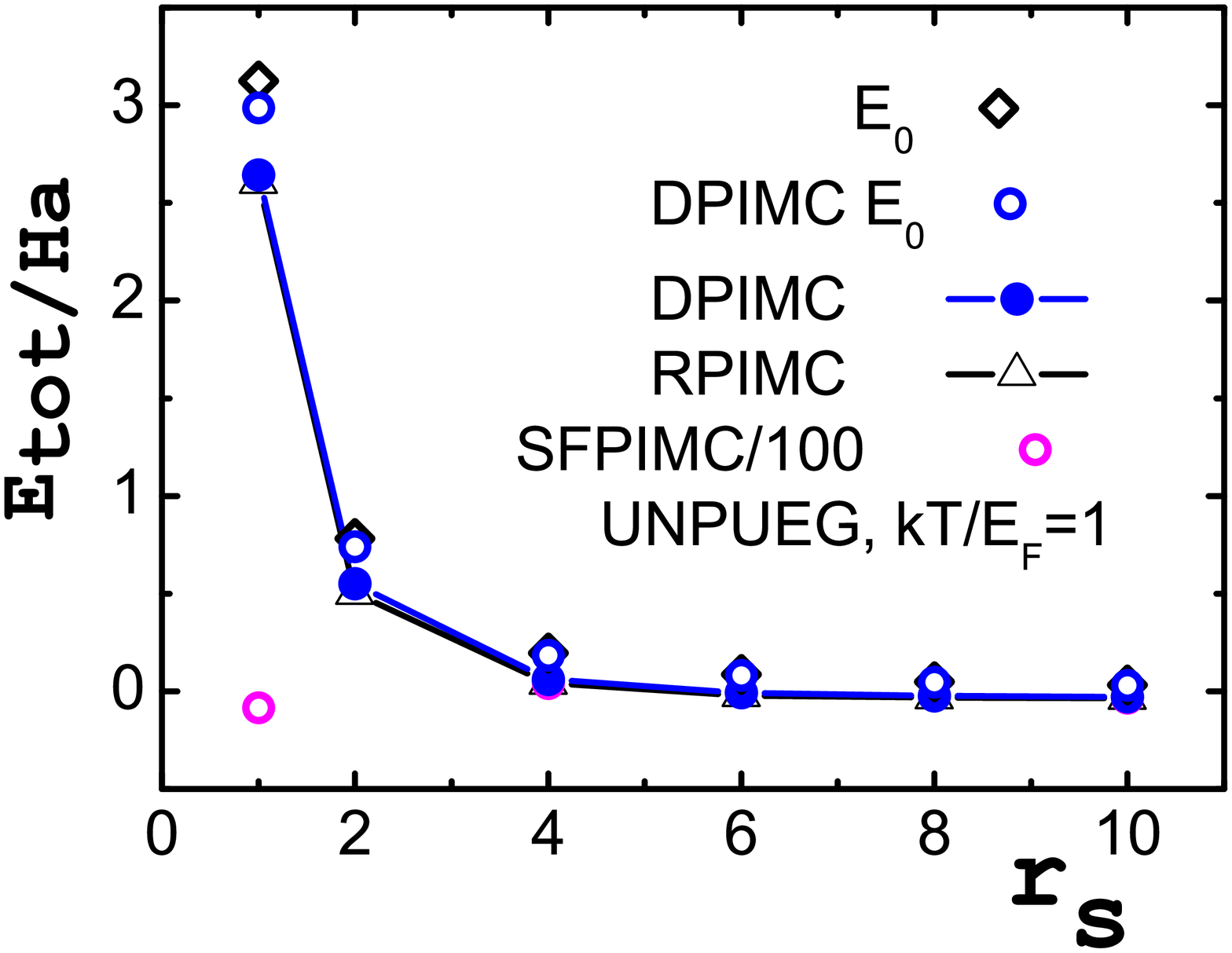}
\includegraphics[width=8.1cm,clip=true]{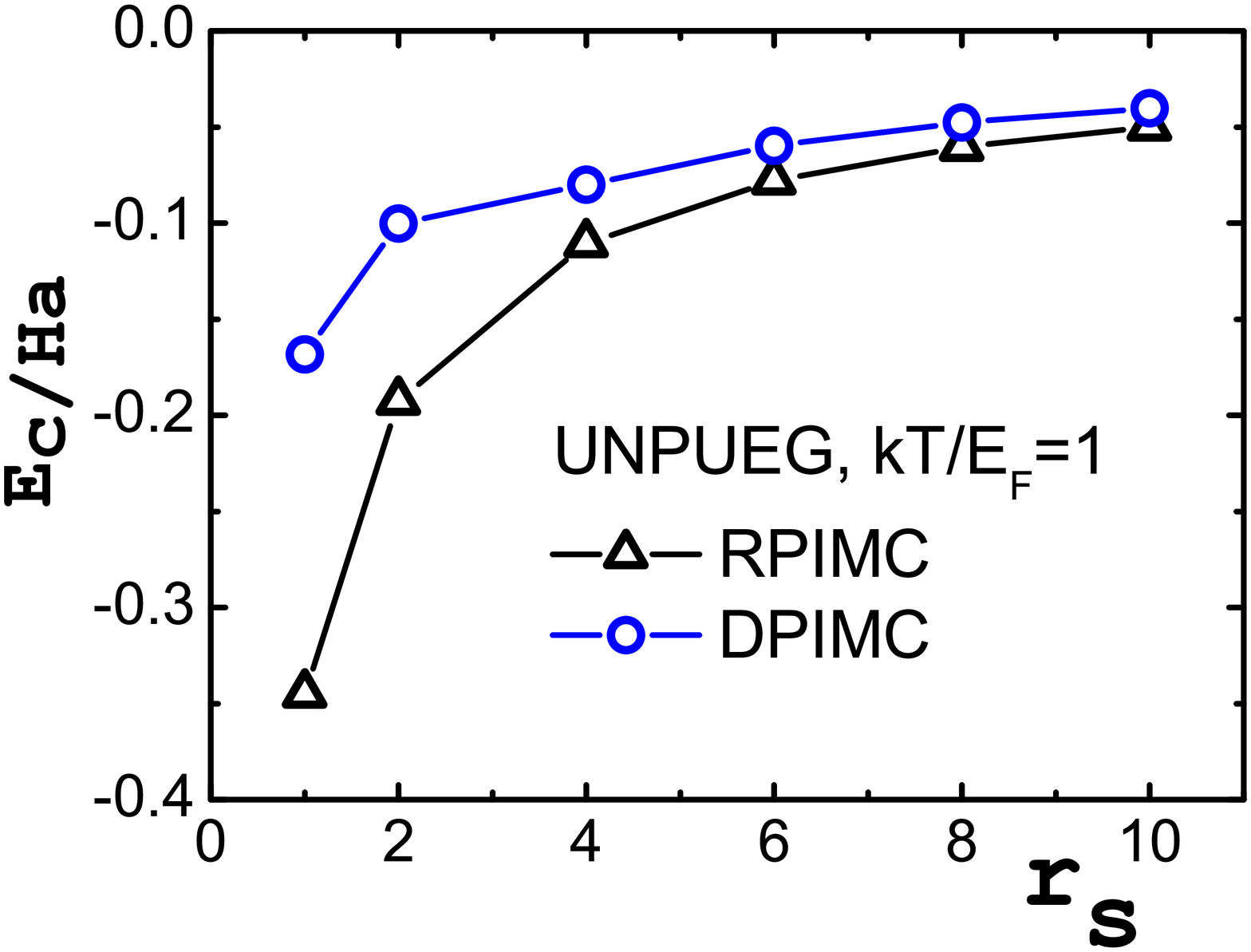}
\includegraphics[width=8.1cm,clip=true]{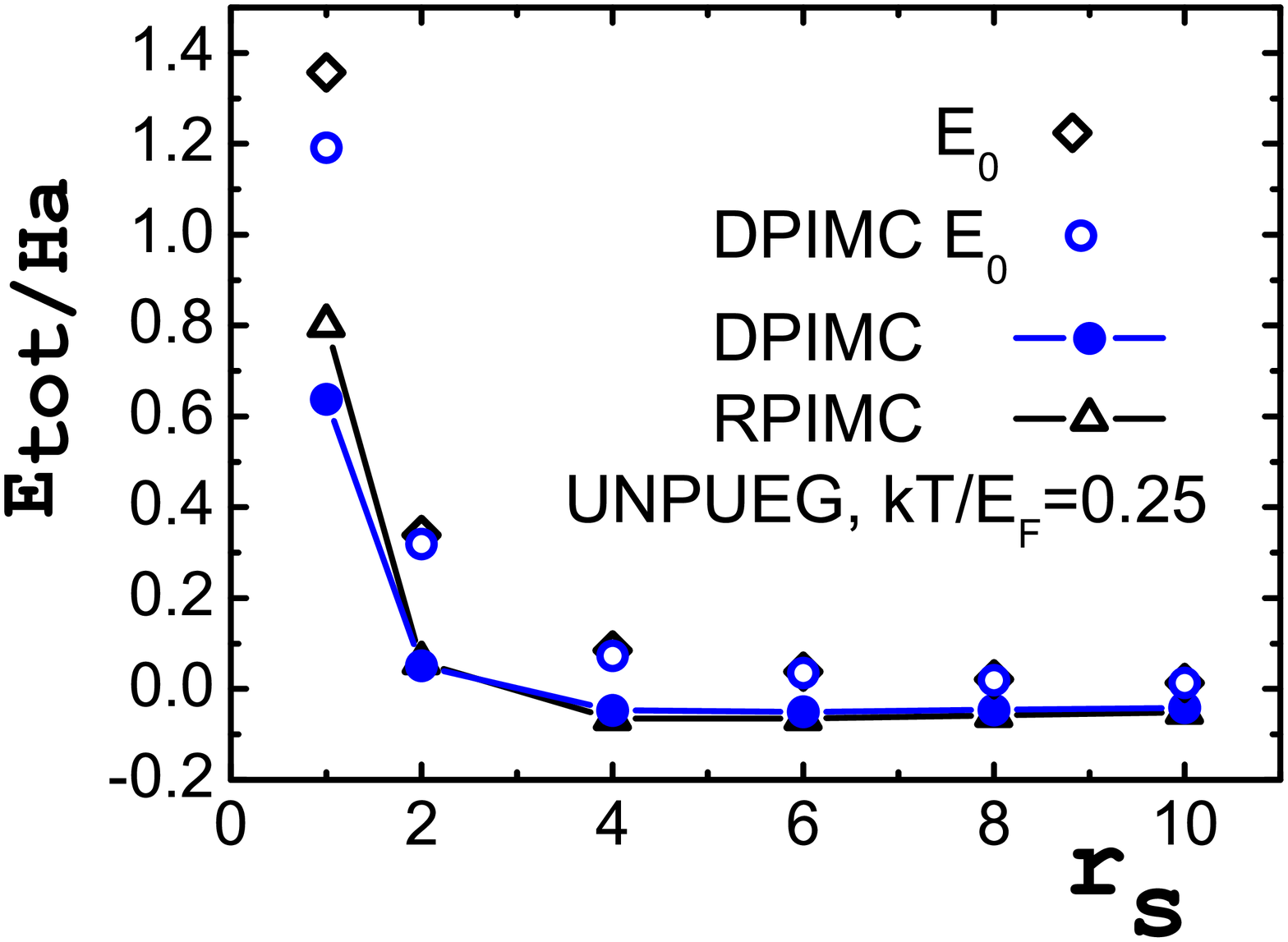}
\includegraphics[width=8.1cm,clip=true]{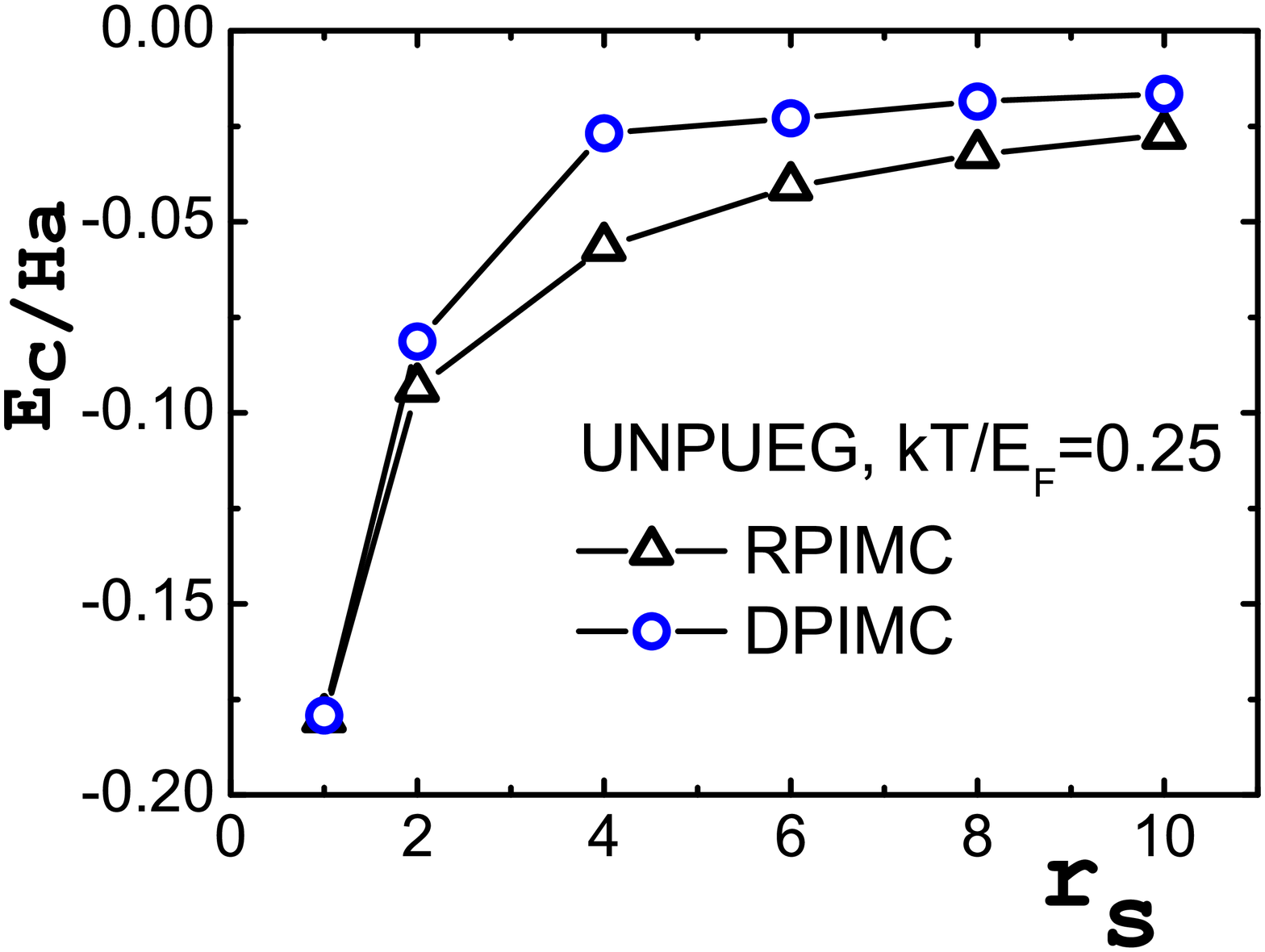}
\includegraphics[width=8.1cm,clip=true]{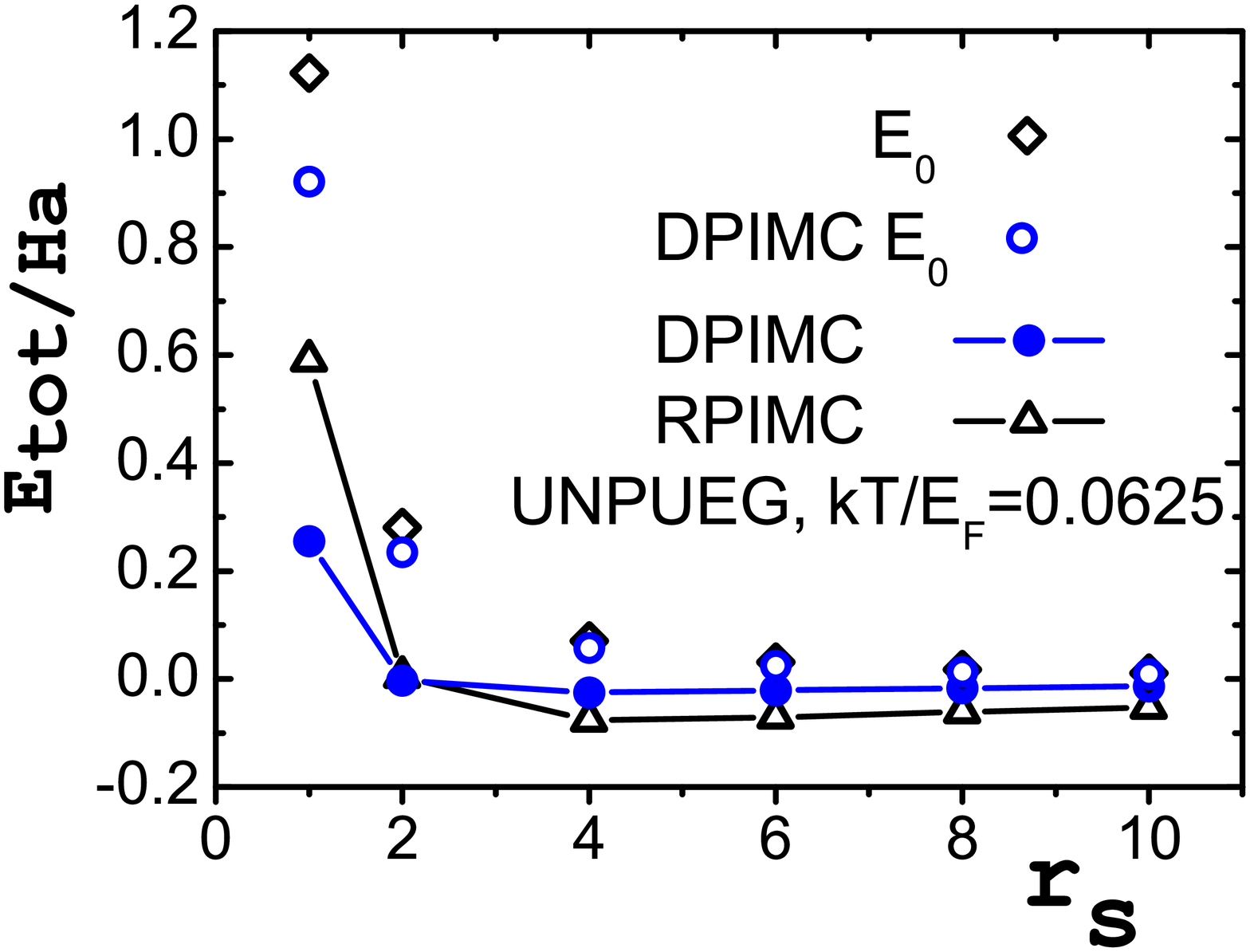}
\includegraphics[width=8.1cm,clip=true]{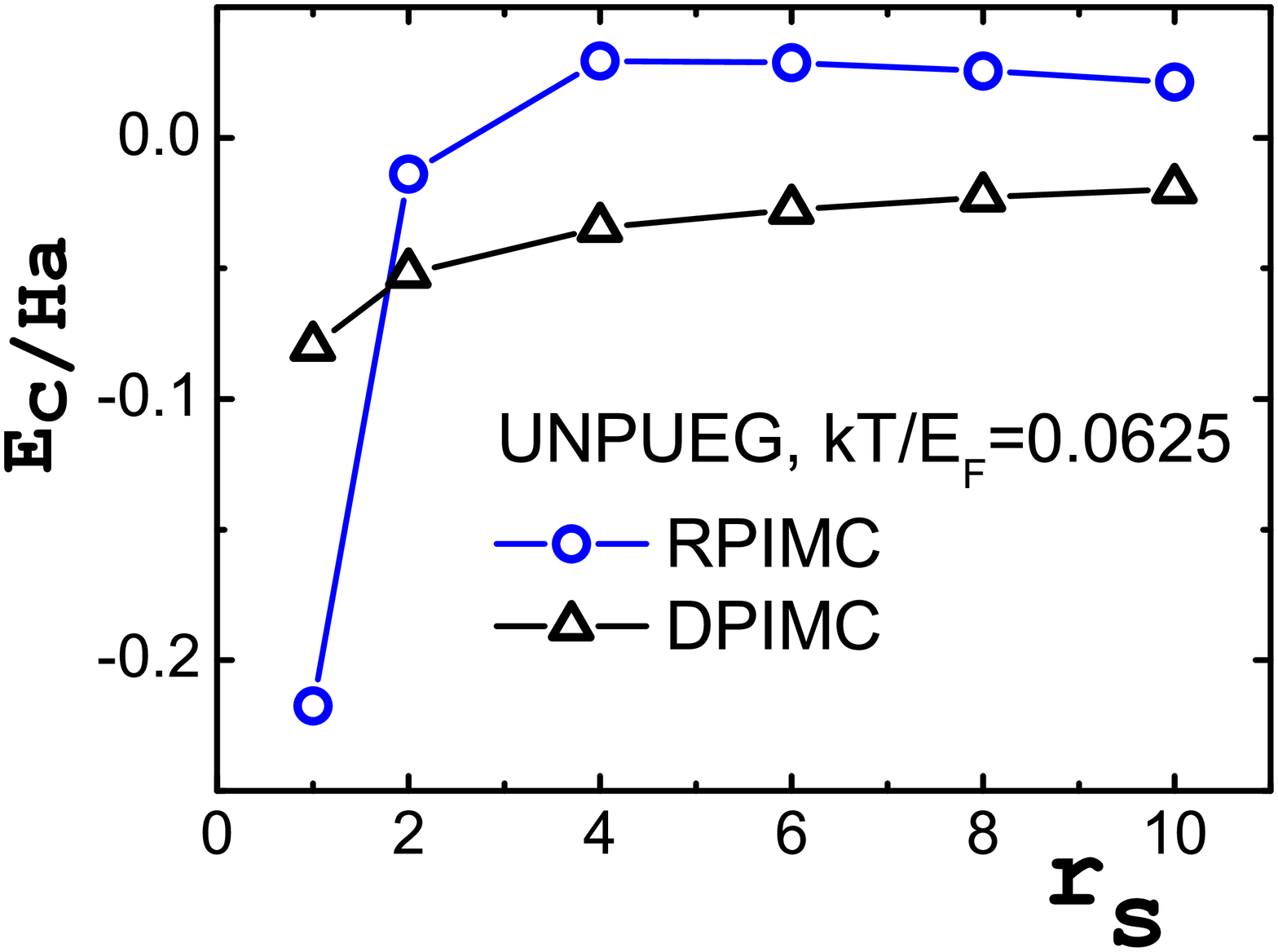}\\
\caption{(Color online) Same as Fig.~\ref{fig:Pol} but for an unpolarized electron gas.
\label{fig:UnPol}
}
\end{figure}



\section{Discussion}\label{summary}
In this paper we have presented new and improved fermionic path integral Monte Carlo simulations for the thermodynamic properties of the uniform electron gas over a wide density and temperature range relevant for warm dense matter conditions. The results were produced with a previously developed code for two-component Coulomb systems where here, for the treatment of the homogeneous background, the interaction energy contributions of the second component have been turned off. The present simulations contain two main improvements: first, we included long-range Coulomb effects via Ewald summation, where an angle average was performed, as proposed by Yakub et al. \cite{Yakub}. Second, we improved the treatment of exchange, in the case of strongly degenerate electrons. For cases where the electron thermal wavelength exceeds the length of the simulation cell, we included not only exchanges between particles in the main Monte Carlo cell but also with electrons from the nearest neighbor cells. This had a drastic effect on the computed energy, and the good accuracy was demonstrated on the limiting case of the energy of an ideal Fermi gas which matches the analytically known results very well.
We oberved deviations for the lowest temperatures ($\Theta=0.0625, 1$) and highest densities, $r_s=1$. Evidently, here it would be desirable to include also exchanges with the next to nearest neighbor cells. These simulations are an order of magnitude more CPU time expansive and will be presented elsewhere.

For the interacting case we compared our direct fermionic PIMC simulations with recent restricted PIMC simulations by Brown et al. \cite{brown13}. There is very good agreement for all densities, $r_s \gtrsim 1$, for temperatures above the Fermi energy. For lower temperatures we observe significant deviations from the results of Ref.~\cite{brown13}. Furthermore, a comparison with the direct fermionic PIMC simulations of the same authors reveales substantial deviations for $r_s \lesssim 4$, indicating that the latter results are not converged with respect to the number $M$ of high-temperature factors (beads). Based on our data for the ideal Fermi gas we conclude that our results are more reliable at high degeneracy than the available RPIMC results due to the improved approach to the fermionic exchange. 
At the same time, to go to lower temperatures or/and higher densities requires further improvements to compensate the rapid lowering of the average sign.
While the total energies in our DPIMC simulations are in overall good agreement with the RPIMC results of Ref.~\cite{brown13} we observe relatively large divations of the correlation energies. This is, in part, attributed the different number of particles in the two simulations and to finite size effects which are included in Ref.~\cite{brown13} but not in our results. Here more analysis has to be done to come to a one to one comparison.

Finally, we mention an alternative approach to access the correlated electron gas at high degeneracy ($r_s<1$). It is provided by the recently developed Configuration path integral Monte Carlo (CPIMC) approach that was successfully applied to finite Fermion systems in traps \cite{schoof_cpp_11}, for an introduction to the method, see Ref.~\cite{simon_springer14}. Comparisons with the CPIMC results for the uniform electron gas \cite{schoof14} will be presented in a forthcoming paper.

\section*{Acknowledgements}
We acknowledge stimulating discussions with T. Schoof (Kiel) and J.W. Dufty and V. Karasiev (University of Florida).
This work has been supported by the Deutsche Forschungsgemeinschaft via SFB TR-24. Computations were performed at the ``Fermion'' compute-cluster of the Institute for Theoretical Physics and Astrophysics of Kiel University and at the North-German Supercomputing Center (HLRN) via grant SHP006.

\end{document}